\documentclass[useAMS,usenatbib]{mn2e}

\usepackage{graphicx,amssymb,amstext} 

\begin{document}

\title[The IIFSCz: luminosity functions, evolution and galaxy bias]{The Imperial IRAS-FSC Redshift Catalogue: luminosity functions, evolution and galaxy bias} 
\author[L. Wang and M. Rowan-Robinson]{
\parbox[t]{\textwidth}{
Lingyu Wang$^{1,2}$\thanks{E-mail: lingyu.wang@sussex.ac.uk}, Michael Rowan-Robinson$^{1}$}
\\
$^{1}$Astrophysics Group, Blackett Laboratory, Imperial College of Science Technology and Medicine, London SW7 2AZ\\
$^{2}$Astronomy Centre, Department of Physics and Astronomy, University of Sussex, Falmer, Brighton, BN1 9QH\\
}

\date{Accepted . Received ; in original form }

\maketitle

\begin{abstract}
We present the luminosity function and selection function of 60 $\mu m$ galaxies selected from the Imperial IRAS-FSC Redshift Catalogue (IIFSCz). Three methods, including the $1/V_{\textrm{max}}$ and the parametric and non-parametric maximum likelihood estimator, are used and results agree well with each other. A density evolution $\propto (1+z)^{3.4\pm0.9}$ or a luminosity evolution $\propto \exp(1.7~t_L / \tau)$ where $t_L$ is the look-back time is detected in the full sample in the redshift range [0.02, 0.1], consistent with previous analyses. Of the four infrared subpopulations, cirrus-type galaxies and M82-type starbursts show similar evolutionary trends, galaxies with significant AGN contributions show stronger positive evolution and Arp 220-type starbursts exhibit strong negative evolution. The dominant subpopulation changes from cirrus-type galaxies to M82-type starbursts at $\log_{10} (L_{60} / L_{\odot}) \approx 10.3$.

In the second half of the paper, we derive the projected two-point spatial correlation function for galaxies of different infrared template type. The mean relative bias between cirrus-type galaxies and M82-type starbursts, which correspond to quiescent galaxies with optically thin interstellar dust and actively star-forming galaxies respectively, is calculated to be $b_{\textrm{cirrus}}/b_{\textrm{M82}}=1.25\pm0.07$. The relation between current star formation rate (SFR) in star-forming galaxies and environment is investigated by looking at the the dependence of clustering on infrared luminosity. We found that M82-type actively star-forming galaxies show stronger clustering as infrared luminosity / SFR increases. The correlation between clustering strength and SFR in the local Universe seems to echo the basic trend seen in star-forming galaxies in the Great Observatories Origins Deep Survey (GOODS) fields (at $z\sim1$).
\end{abstract}


\section{Introduction}
The study of galaxy distribution on large scales ($\sim10 ~h^{-1}$ Mpc) where density fluctuations are still in the linear regime ($\delta \rho / \rho \ll 1$) has been an important tool in constraining cosmological models. However, the unknown relationship between the distribution of dark matter and that of luminous objects, usually referred to as galaxy bias, has complicated the extraction of cosmological parameters, especially when we compare results from surveys of different galaxy subpopulations or of varying depths. At small scales ($\sim1 ~h^{-1}$ Mpc) where nonlinear effects become important, detailed studies of galaxy bias can improve our understanding of galaxy formation and evolution which is still unclear due to the convolution of gravitational evolution and complex astrophysical processes (e.g. gas cooling, fragmentation, star formation, feedback). This paper attempts to measure the relative galaxy bias over the intermediate scale ($1\sim10$) Mpc, i.e. the weakly nonlinear scale, as we are constrained by low galaxy density at small scales.     

Several models of galaxy bias have been developed over the past 20 years or so. The simplest one is to assume galaxies trace the underlying mass in a linear, deterministic way, $\delta_{\textrm{g}} = b \delta_{\textrm{m}}$ where $\delta_{\textrm{g}}$ and $\delta_{\textrm{m}}$ is the galaxy and matter density perturbation (or contrast) field and $b$ is the linear bias parameter. In the peaks bias model (Kaiser 1984; Bardeen et al. 1986), the relationship between the correlation function of galaxies and that of the underlying dark matter satisfies $\xi_{\textrm{g}}(r)=b^2 \xi_{\textrm{m}}(r)$, based on statistics of the density peaks in a Gaussian random field. On large scales, it seems reasonable to assume a scale independent linear galaxy bias if galaxies form in virialized dark matter halos. On intermediate and small scales, the situation is more complicated. In more realistic galaxy bias models, $b$ is thought to be time-varying, scale-dependent, non-local, or even stochastic to some extent. N-body simulations of dark matter and semi-analytic models of galaxy formation predict a scale-dependent relation between galaxies and dark matter due to the mass-dependence of galaxy formation efficiency as well as a luminosity-dependent bias for galaxies with luminosities above $L^*$ (Benson et al. 2000; Benson et al. 2001). Attempts to understand galaxy bias have also been made in the halo occupation distribution (HOD) framework where the relationship between the galaxy distribution and the underlying dark matter is encoded in the conditional probability function $P(N|M)$ which gives the number of galaxies of a given type in a halo of virial mass $M$ (Jing et al. 1998; Ma \& Fry 2000; Peacock \& Smith 2000; Cooray \& Sheth 2002; Zehavi et al. 2005; Tinker et al. 2005; Zheng, Coil \& Zehavi 2007; Simon et al. 2008).  
 
Observationally, several manifestations of galaxy bias are known. From dipole analysis, different values of the $\beta \equiv \Omega_m^{0.6}/b$ parameter have been reported using different types of mass tracer (Rowan-Robinson et al. 2000; Kocevski, Mullis \& Ebeling 2004; Erdo\u{g}du et al. 2006 and references therein). The well-known morphology-density relation (Dressler 1980; Goto et al. 2003) states that late-type galaxies (disk-dominated) prefer low density regions and early-type (bulge-dominated) galaxies dominate in dense cluster cores. There is also a correlation between colour and morphology (the majority of blue galaxies are late-type galaxies, while red galaxies are mostly early-type galaxies), as well as a correlation between colour and luminosity (red galaxies are systematically more luminous than blue galaxies). Therefore, galaxy clustering is expected to be dependent on physical properties such as morphology, colour, surface brightness and luminosity. Using a total of over $100,000$ galaxies from the 2dF Galaxy Redshift Survey (2dFGRS), Norberg et al. (2001) found a scale-independent luminosity dependence of galaxy bias of the form $b/b^*=0.85+0.15L/L^*$ which means the luminosity-dependence is conspicuous above $L^*$. It agrees with the conclusions of Benoist et al. (1996) based upon 3,600 galaxies selected from the Southern Sky Redshift Survey 2 at the faint end, but it exhibits a less steep rise at the bright end. Zehavi et al. (2002, 2005) studied galaxy clustering in the SDSS, which not only confirmed the conclusions of Norberg et al. (2001) but also probed the relative bias behaviour to fainter galaxies. Madgwick et al. (2003) divided their sample from the 2dFGRS into relatively passive and active star-forming galaxy subsamples. A scale-dependent relative bias was found in comparing the clustering strength of the two subsamples. On scales smaller than 8 $h^{-1}$ Mpc, the relative bias between the two spectral classes is $b_{\textrm{passive}}/b_{\textrm{active}}=1.45\pm0.14$. On scales larger than 10 $h^{-1}$ Mpc, the relative bias reduces to unity. Zehavi et al. (2002; 2005) used a luminosity-dependent colour-cut to divide their volume-limited sample from the SDSS. The colour-dependence of the correlation function is found to be very similar to that found in Madgwick et al. (2003). In addition, the luminosity-dependence for red galaxies is such that both faint and luminous red galaxies cluster more strongly than moderately bright red galaxies around $L*$ while a luminosity-dependence for blue galaxies is not noticeable (Wang et al. 2007; Swanson et al. 2008; Cresswell \& Percival 2009). 

The optical colour of a galaxy is connected with a complicated combination of star formation history and current star formation rate (SFR). In contrast, it is well known that the far-infrared emission is a good SFR indicator due to the re-processing of starlight by dust. So in principle one can measure clustering strength of infrared galaxies to learn about the relation between environment and SFR. So far, little is known about the clustering of infrared galaxies and its dependence on various physical parameters. The largest sample of local infrared galaxies is provided by IRAS. The correlation length of IRAS galaxies ($r_0\sim4 ~h^{-1}$ Mpc) is similar to that of local star-forming/blue galaxies found in the SDSS and 2dFGRS (Saunders et al. 1992; Fisher et al. 1994). Clustering measurements of infrared galaxies did not show significant dependence on infrared luminosity (Beisbart \& Kerscher 2000; Szapudi et al. 2000; Hawkins et al. 2001). This was thought to be a result of the combined effect of the dark matter halo mass and SFR in determining the far-infrared luminosity. From the SFR - density relation, we know that star-forming galaxies preferentially reside in low density environments (G\'{o}mez et al. 2003). On the other hand, the gas content in a galaxy is largely dependent on the mass of the host dark matter halo. The dependence of clustering on infrared galaxy types is not entirely clear either. With a few thousand galaxies in the QDOT IRAS galaxy redshift survey (Lawrence et al. 1999), Mann, Saunders \& Taylor (1996) did not detect any significant variation in the clustering of warm and cool IRAS galaxies separated by the far-infrared emission temperature (above or below 36 K respectively), assuming the far-infrared spectrum can be fitted by $S_{\nu}\propto \nu B_{\nu}(T)$. Hawkins et al. (2001), using the much bigger IRAS PSCz sample (Saunders et al. 2000), found that the relative bias between cool and warm infrared galaxies is around 1.5 on scales between 1 and 10 $h^{-1}$ Mpc. 

Our aim is to investigate the relation between star formation activity in star-forming galaxies and environment using infrared samples in the general field in the local Universe. We study the clustering dependence on infrared luminosity and infrared template type, taking advantage of a new redshift catalogue of over 60,000 galaxies selected at $60~\mu m$ from the IRAS Faint Source Catalog (FSC). The remainder of this paper is organised as follows. Section~\ref{sample} briefly describes the galaxy sample used. Details of the new catalogue can be found in Wang \& Rowan-Robinson (2009). In Section~\ref{LF and SF}, we derive the luminosity function and the selection function using various methods which will then be used to construct random catalogues in measuring the clustering signal with flux-limited samples. In Section~\ref{galaxybias}, clustering dependence on infrared template type and infrared luminosity is investigated using the projected two-point spatial correlation function. Finally, discussions and conclusions are presented in Section~\ref{discussions and conclusions}. Throughout the paper, unless otherwise stated, we assume $H_0=72~\textrm{km}~\textrm{s}^{-1}~\textrm{Mpc}^{-1}$ and a flat universe with $\Omega_M=0.3$ and $\Omega_{\Lambda}=0.7$. Luminosities are quoted in units of $h_{72}^{-2}~L_{\odot}$.

\section{The galaxy sample}
\label{sample}

\begin{figure}\centering
\includegraphics[height=3.0in,width=3.4in]{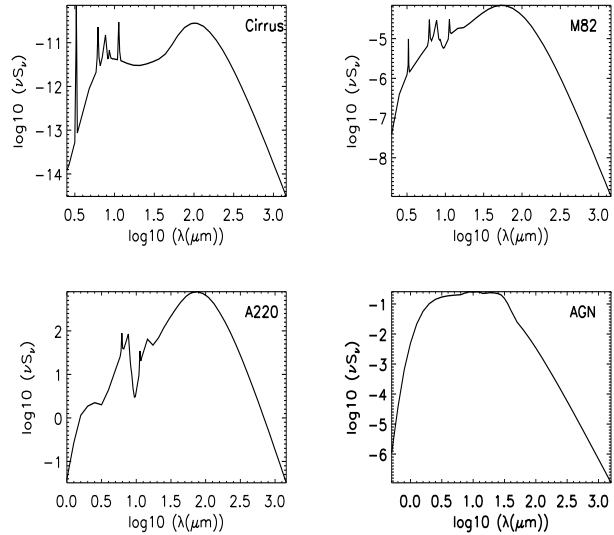}
\caption{The four infrared templates used to fit IRAS fluxes from 12 to 100 $\mu$m for galaxies in the IIFSCz.}
\label{fig:infrared-type}
\end{figure}

The Imperial IRAS-FSC Redshift Catalogue (IIFSCz; Wang \& Rowan-Robinson 2008) is a recently constructed catalogue containing 60,303 galaxies selected at $60~\mu m$ from the IRAS FSC. The IIFSCz covers most of the sky at $|b|>20^{\circ}$ ($\sim61\%$ of the entire sky). $55\%$ of the sources have spectroscopic redshifts obtained from past redshift surveys and a further $20\%$ have optical, near-infrared and/or radio identifications for which photometric redshifts were estimated through the training set or template-fitting method. We also determine the best-fit infrared template for each galaxy with either spectroscopic or photometric redshift. Fig.~\ref{fig:infrared-type} shows the main four infrared template types, infrared cirrus (dust emission due to the interstellar radiation field), M82 starburst (interaction-induced starburst), AGN dust torus emission and Arp 220 starburst (merger-induced starburst). 

The intrinsic $90\%$ completeness limit of the IRAS FSC is at around $F_{60}=0.36$ Jy. In order to reduce incompleteness-induced problems, we have selected a sample of around 22,500 galaxies with $F_{60}\ge0.36$ Jy. At this flux limit, around $75\%$ of our sample have spectroscopic redshifts, $15\%$ have photometric redshifts and $10\%$ do not have any redshift estimates. Almost all galaxies in areas covered by the SDSS DR6 survey have redshift, either spectroscopic or photometric. Thus, the redshift completeness is a spatially varying function and this is taken into account in this paper. 

\begin{figure}\centering
\includegraphics[height=3.0in,width=3.4in]{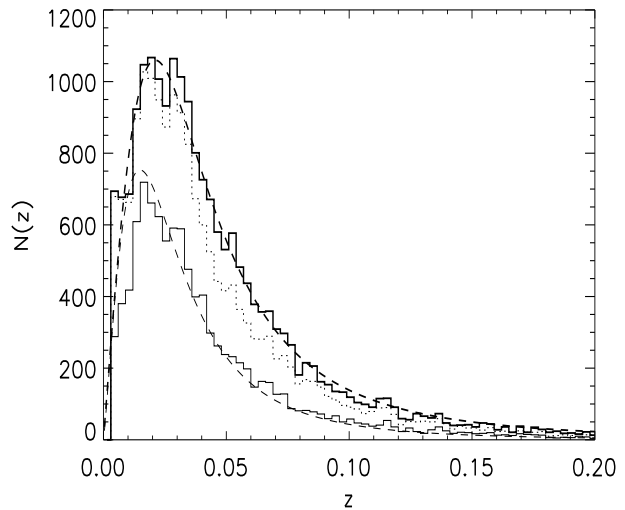}
\caption{The redshift distribution of the IRAS PSCz galaxies (thin solid line) and IIFSCz galaxies with either spectroscopic or photometric redshifts (thick solid line) or IIFSCz galaxies with spectroscopic redshifts only (dotted line). The dashed curves are obtained from the selection functions of the PSCz (thin dashed curve) and IIFSCz (thick dashed curve). Galaxies with redshifts $z<0.003$ are excluded.}
\label{fig:z_hist}
\end{figure}

\begin{figure}\centering
\includegraphics[height=3.0in,width=3.4in]{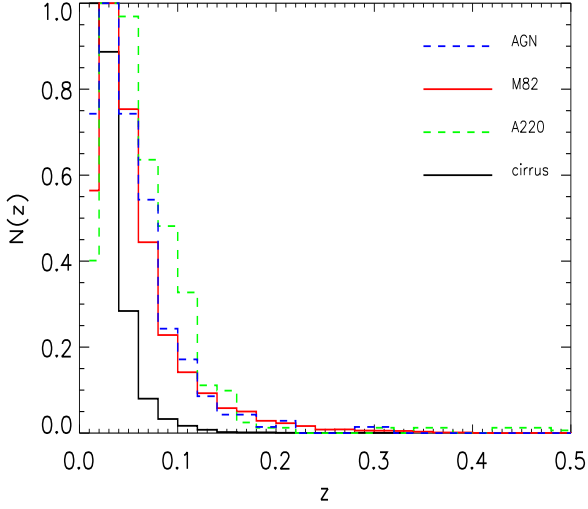}
\caption{The redshift distribution of each infrared population in the IIFSCz. The peaks are normalised to 1.}
\label{fig:zhist-type}
\end{figure}

We convert heliocentric redshift to redshift in the Local Group frame. Redshifts are used as distance indicators without corrections for peculiar motions. In consequence, we do not include galaxies below redshift $z=0.003$ for which $z$ is a poor estimate of the distance, in our sample. In Fig.~\ref{fig:z_hist}, the redshift distribution of galaxies in our sample is plotted in comparison with galaxies in the $IRAS$ PSCz. Fig.~\ref{fig:zhist-type} shows the normalised redshift distribution for each infrared galaxy type in our sample. 

Cirrus-type galaxies are mainly found in the local Universe ($z<0.06$), while the other three types are more common at relatively high redshift. So, cirrus-type galaxies are almost absent at the high luminosity end. This transition in the predominant subpopulation in the mixture across the redshift (or luminosity) range is important in understanding the clustering dependence on luminosity for the full sample. We will return to this point in Section 4.

\section{The luminosity and selection function}
\label{LF and SF}

The selection function $S(z)$ (hereafter SF) of a flux-limited survey is defined as the mean comoving number density of galaxies that can be observed at redshift $z$. It is the integration of the luminosity function $\Phi (L)$ (hereafter LF) at redshift $z$ which specifies the luminosity distribution of the galaxy population observed in the survey,
\begin{equation}
S(z)=\int_{L_{\textrm{min}(z)}}^{\infty} \Phi_z (L)dL,
\label{SFdefinition}
\end{equation}
where $L_{\textrm{min}(z)}$ is the minimum luminosity a galaxy at redshift $z$ can have in order to remain above the flux limit. The inherent assumption is that the LF does not depend on the local density field. However, it has been known for some time that the LF is dependent on galaxy type which is in turn related to the environment. Therefore, the assumption that the bivariate distribution function of luminosity and position is separable is not strictly valid.

\subsection{Methods}

The LF is of fundamental importance in constraining models of galaxy formation and evolution. Various techniques have been proposed to measure it, such as the maximum-volume estimator $1/V_{\textrm{max}}$ (Schmidt 1968) and the parametric (Sandage, Tammann \& Yahil 1979, hereafter STY) and non-parametric maximum likelihood estimator (Efstathiou, Ellis \& Peterson 1988; Springel \& White 1998, hereafter SW98). The $1/V_{\textrm{max}}$ method assumes a uniform number density throughout the observed volume and therefore is vulnerable to density inhomogeneities present in a survey. The maximum likelihood estimators have the advantage of being insensitive to any local clustering effect, but lose information on the overall normalisation. 

The STY method requires a prior knowledge of the appropriate functional form of the LF. The probability of a source at redshift $z_i$ having luminosity in the range $[L_i,L_i+dL]$ is 
\begin{equation}
p(L_i|z_i)=\frac{ n(z_i, L_i) }{ \int_{L_{\textrm{min}(z_i)}}^\infty n (z_i, L')dL' }=\frac{\Phi_{z_i} (L_i)}{S(z_i)}.
\end{equation}
In Bayesian statistics, the posterior probability of the parameters characterising the LF given the data is proportional to the likelihood function (i.e. the probability of the data given the parameters) assuming a uniform prior probability of the parameters. Therefore, the best-fit parameters in the LF are the ones which maximise the likelihood given by
\begin{equation}
\mathcal{L}=\prod_{i} p(L_i|z_i).
\end{equation}

Saunders et al. (1990; hereafter S90), using the STY method, determined the present-epoch 60\,$\mu$m LF from 2818 IRAS galaxies which is a Gaussian combined with a power law,
\begin{equation}
\varphi(L)=C\left(\frac{L}{L_*}\right)^{1-\alpha} \exp\left[-\frac{1}{2\sigma^2}\log_{10}^2 \left(1+\frac{L}{L_*}\right)\right], 
\label{eqn:Saunders-LF}
\end{equation}
where $C=2.6\times 10^{-2}~h^3~\textrm{Mpc}^{-3}$, $\alpha=1.09$, $\sigma=0.724$, $L_*=10^{8.47}~h^{-2}~L_{\odot}$. Note that $\varphi(L)$ is defined as the LF per decade in luminosity. The normalisation $C$ is derived from the expected number density of sources with $F_{60}\ge0.6$ Jy per unit solid angle.

The parametric maximum likelihood technique does not have an intrinsic measure of goodness-of-fit. Therefore, a non-parametric approach is needed to justify the functional form of the LF or the closely related SF. SW98 proposed a non-parametric maximum likelihood estimator which adopts stepwise power laws to characterise the SF, 
\begin{equation}
S(z)=S_k \left(\frac{z}{x_k}\right)^{m_k},
\label{SFpowerlaw}
\end{equation}
where $k$ is the redshift bin number. For redshift in the $k$ bin ($x_{k-1} < z \le x_k$), $m_k$ is the logarithmic slope of the power law. 

\begin{figure}\centering
\includegraphics[height=3.0in,width=3.4in]{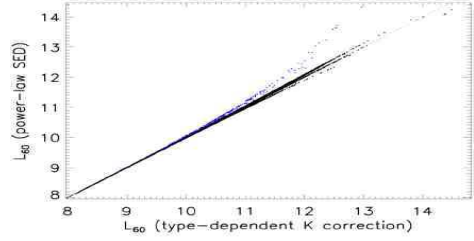}
\caption{The x-axis is the 60 $\mu$m luminosity with a type-dependent K-correction derived from the best-fit infrared template. The y-axis is the 60 $\mu$m luminosity with a K-correction derived from a power-law SED $S_{\nu}\propto \nu^{-2}$. The dotted line corresponds the perfect agreement between the two. The blue dots are AGN dust tori and the black dots are galaxies of any other type in the redshift range $0.003\le z\le4.0$.}
\label{fig:lum1-vs-lum2}
\end{figure}

If we assume a pure density evolution of the form
\begin{equation} 
\bar{n}(z)=\bar{n}_0 g(z)=\bar{n}_0 (1+z)^P,
\label{densityEvolution} 
\end{equation}
the best estimates of $m_k$ can be derived from the following equation,
\begin{equation}
\frac{\partial (\ln \mathcal{L})}{\partial m_k}=\sum_i \frac{\delta_{k,a_i}}{m_k - z_i^m\frac{g'(z_i^m)}{g(z_i^m)}} + T_k=0.
\end{equation}
where
\begin{equation}
T_k = \sum_i\left(\sum_{j=b_i+1}^{a_i} \delta_{j,k} \ln{\frac{x_j}{x_{j-1}}}+\delta_{k,a_i} \ln \frac{z_i^m}{x_{a_i}} - \delta_{k, b_i} \ln \frac{z_i}{x_{b_i}}\right).
\end{equation}
The variance of $m_k$ has the simple form
\begin{equation}
V[m_k] = \left\{ \sum_i \delta_{k,a_i} \left[m_k - z_i^m\frac{g'(z_i^m)}{g(z_i^m)}\right]^{-2} \right\}^{-1}.
\end{equation}
Note that $z_i^m$ is the maximal redshift a galaxy can have such that $L_{\textrm{min}}(z_i^m)=L_i$. Without including luminosity evolution, $z_i^m$ can be derived by solving the following nonlinear equation,  
\begin{equation}
r_i^m = r_i \sqrt{\frac{F_i}{F_{\textrm{min}}}\frac{1+z_i}{1+z_i^m} \Psi \left(\frac{1+z_i^m}{1+z_i}\right)}
\end{equation}
The comoving distance in a $\Lambda$CDM cosmology can be calculated from
\begin{equation}
r_i = \frac{c}{H_0} \int_0^{z_i} \frac{dz'}{\sqrt{\Omega_M (1+z')^3 +\Omega_{\Lambda}}}.
\end{equation}
Here $\Psi$ denotes the K-correction
\begin{equation}
\Psi(\eta)=\frac{ \int R (\nu) f_{\nu} (\nu \eta) d\nu }{\int R(\nu') f_{\nu}(\nu') d\nu'}, 
\end{equation}
where $R(\nu)$ is the responsivity of the photometric band. If we take into account luminosity evolution of the form
\begin{equation}
L_z = L_0 \exp{(Qt_L/\tau)},
\end{equation}
where $\tau$ is the Hubble time and $t_L$ is the lookback time
\begin{equation}
t_L = \tau \int_0^z \frac{dz'}{(1+z')\sqrt{\Omega_M (1+z')^3 + \Omega_{\Lambda}}},
\end{equation}
$z_i^m$ needs to be recalculated accordingly. 

Most of the previous estimates in the literature have derived the K-correction term by assuming a power-law spectrum of the form $S_\nu \propto \nu^{-2}$ (although in some papers, people have tried other methods such as a grey-body model constrained by the 60-100 $\mu$m flux ratio). In light of Fig.~\ref{fig:infrared-type}, this assumption may not seem very problematic over a narrow redshift range except for AGN dust tori. Fig.~\ref{fig:lum1-vs-lum2} compares the K-corrected 60 $\mu$m luminosity derived from the best-fit infrared template\footnote{For galaxies which are detected at 60 $\mu$m only (around $13\%$ of the whole sample), we have assumed a power-law spectrum of the form $S_\nu \propto \nu^{-2}$.} for each galaxy with that derived from a $S_\nu \propto \nu^{-2}$ spectrum over the redshift range $0.003\le z\le4.0$, without evolution correction. As expected, the difference between the two luminosity estimates is small except for AGNs at relatively high redshifts. 

\begin{figure}\centering
\includegraphics[height=3.0in,width=3.4in]{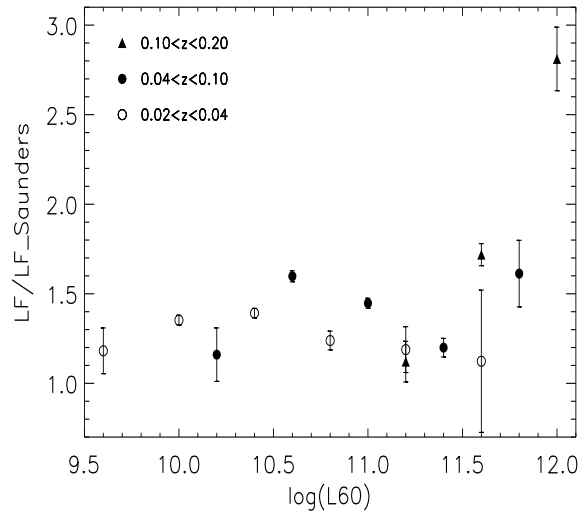}
\caption{The $1/V_{\textrm{max}}$ 60\, $\mu$m LF derived for the full sample in 3 redshift bins without evolution correction. To increase the contrast between LFs in different redshift bins, we have normalised the LF by Saunders et al.'s estimate of the present-day LF, i.e. Eq. 4.}
\label{fig:zbin_all}
\end{figure}

\begin{figure*}\centering
\includegraphics[height=2.8in,width=3.4in]{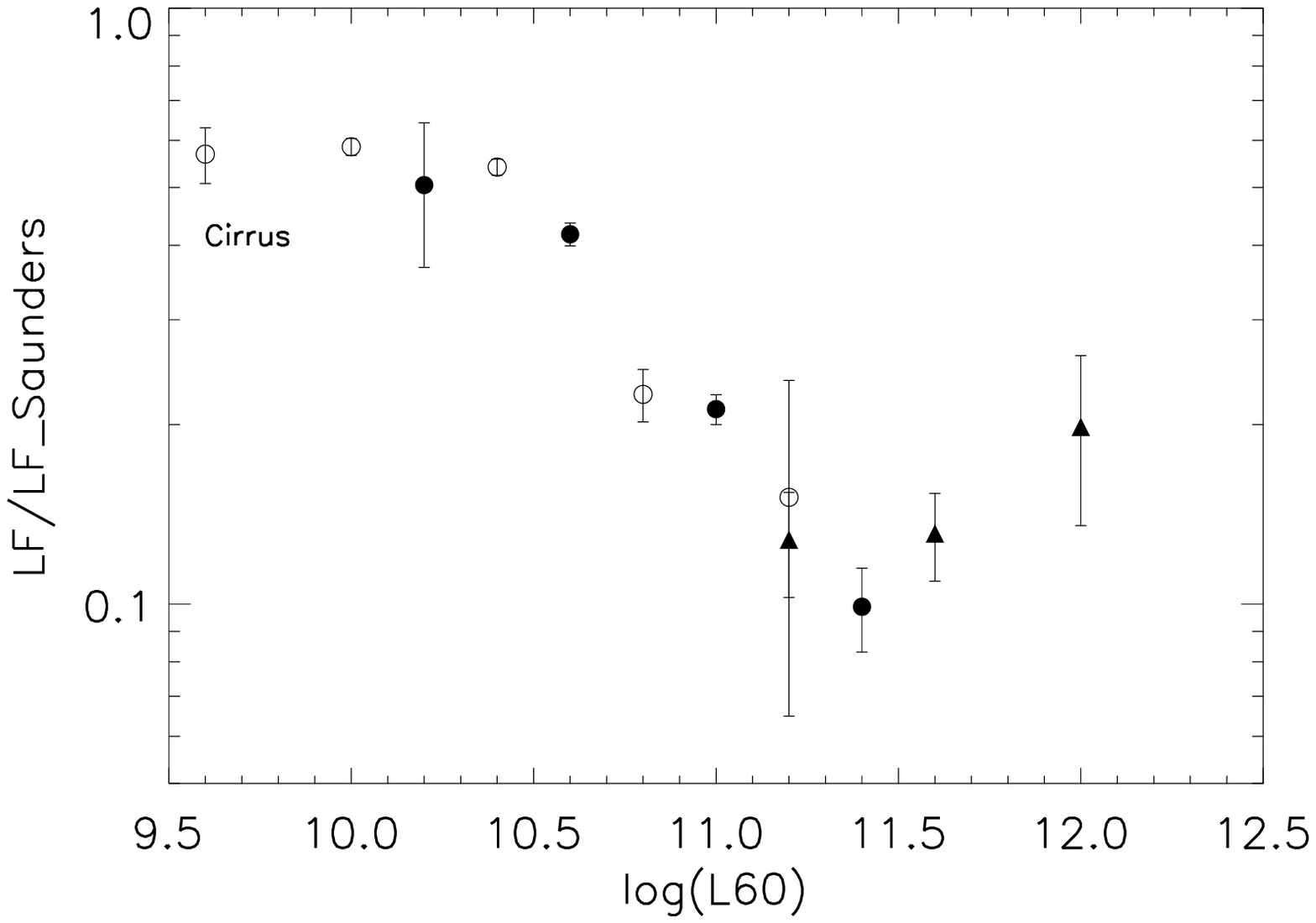}
\includegraphics[height=2.8in,width=3.4in]{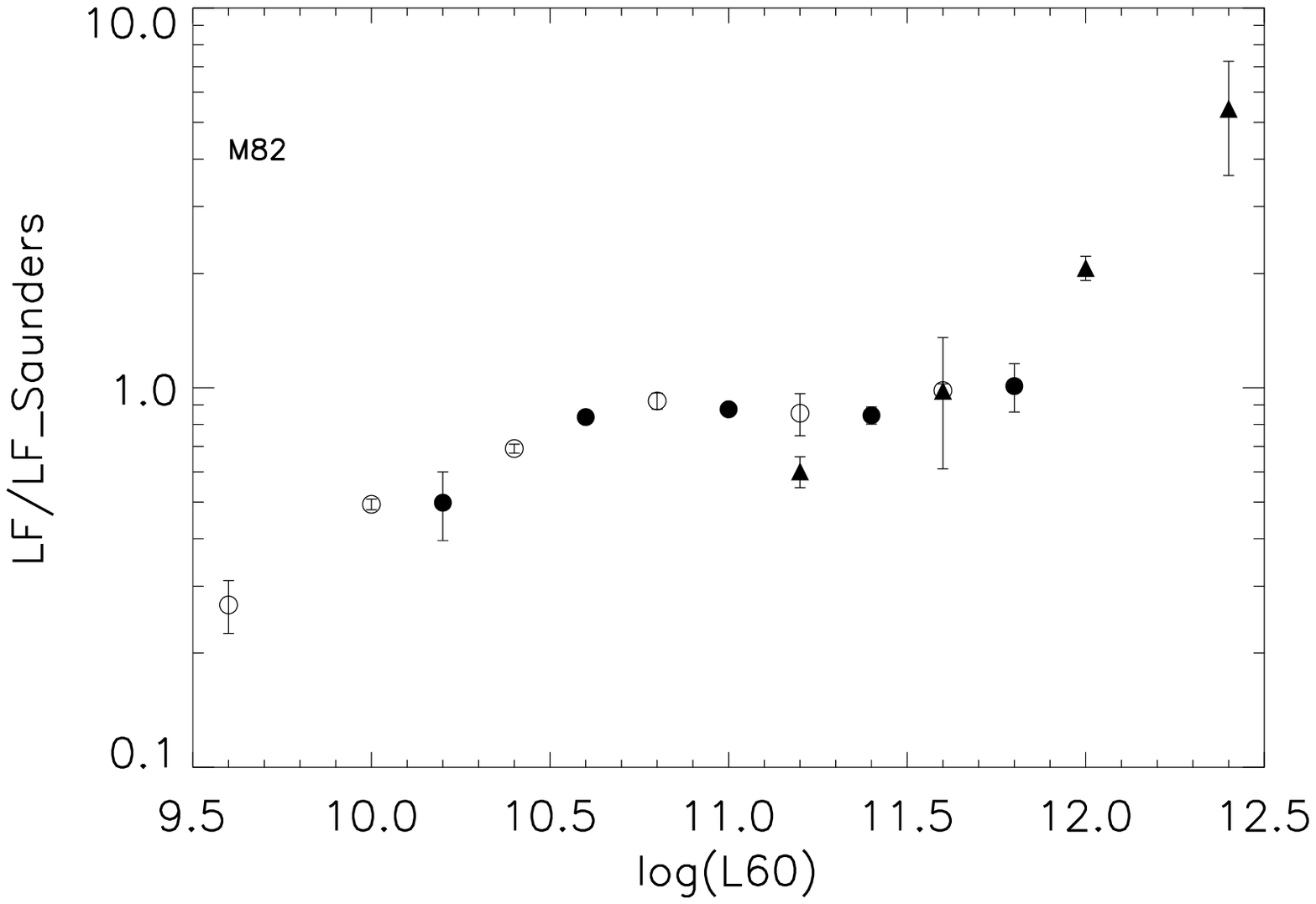}
\includegraphics[height=2.8in,width=3.4in]{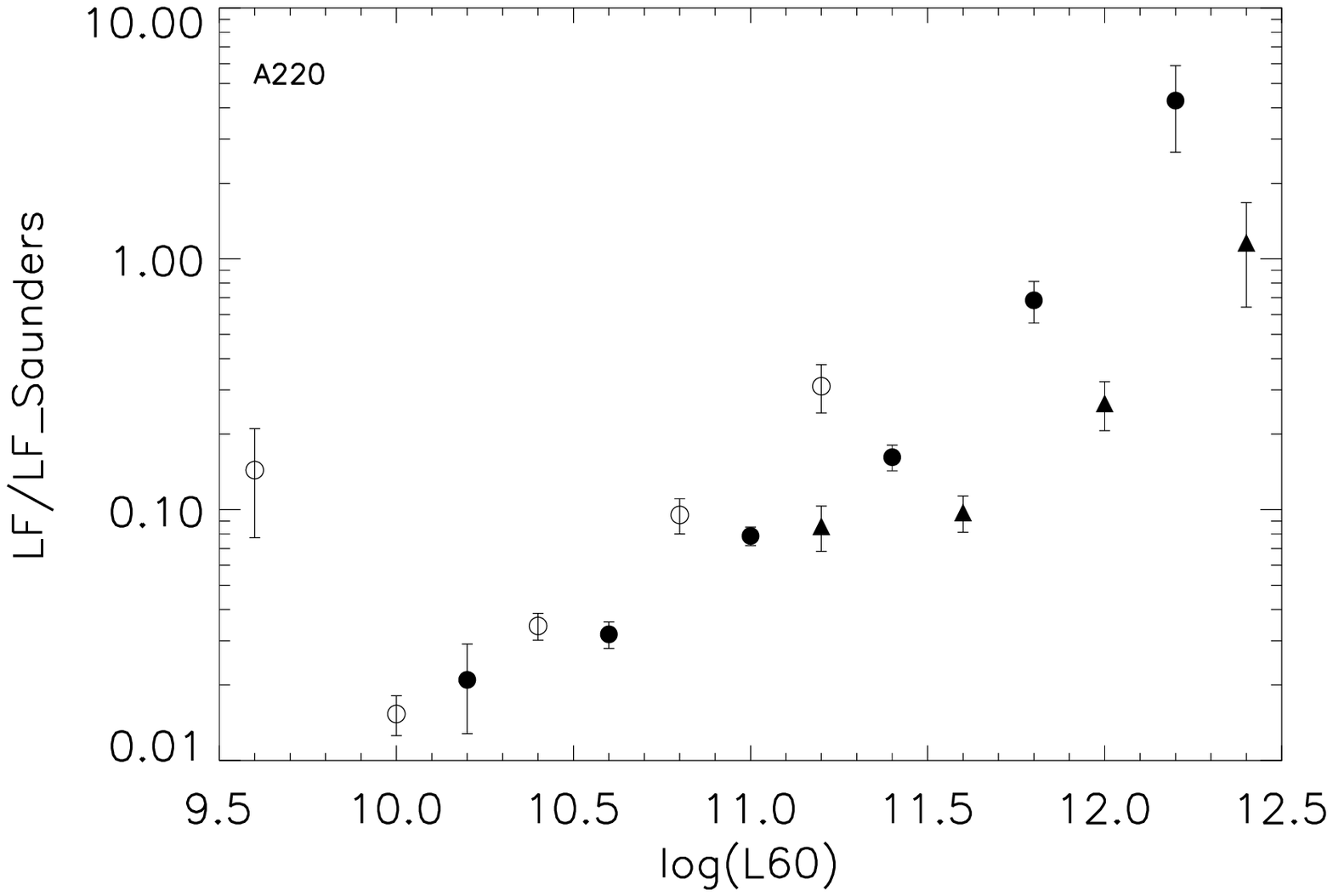}
\includegraphics[height=2.8in,width=3.4in]{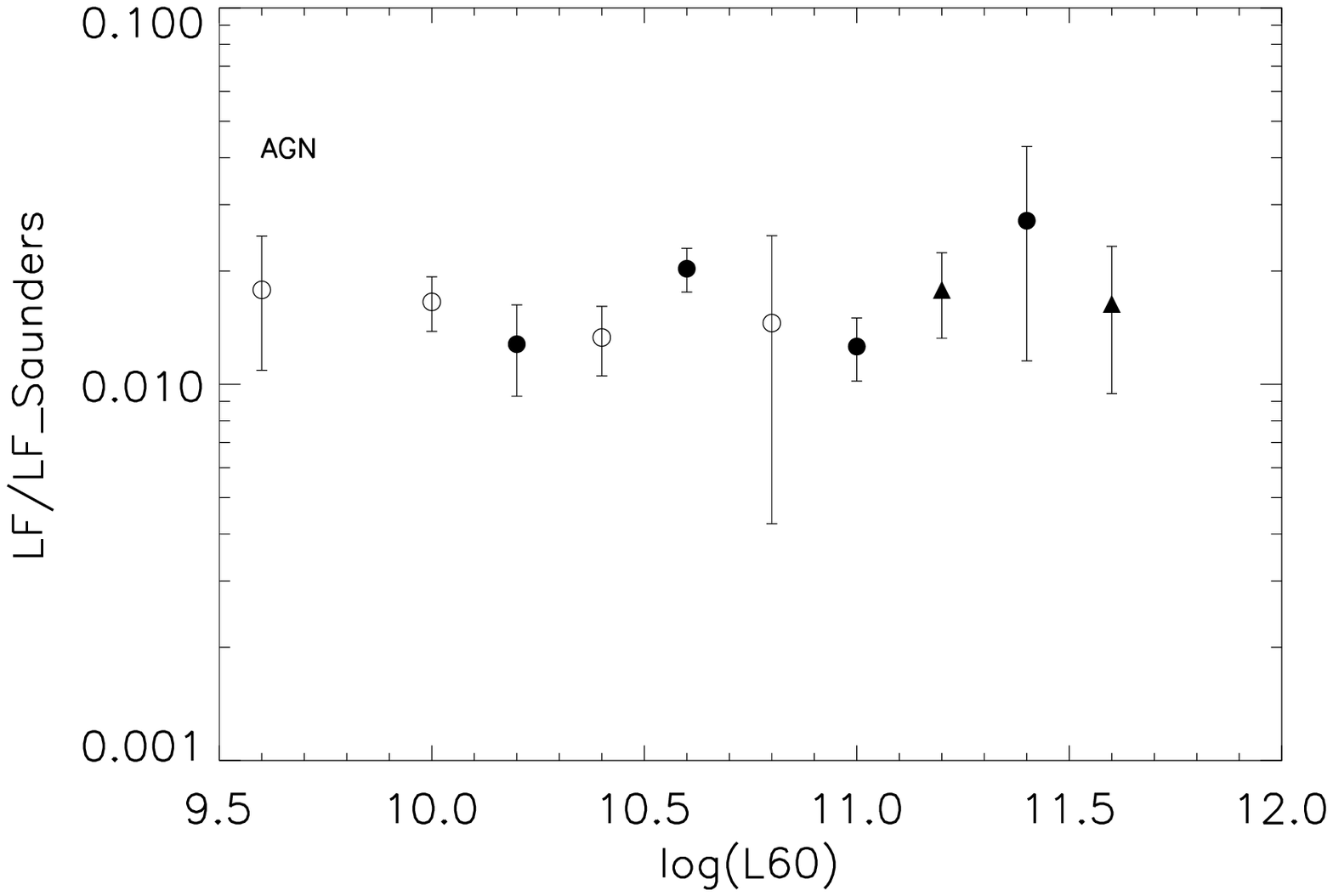}
\caption{The $1/V_{\textrm{max}}$ 60\, $\mu$m LFs derived for each infrared subpopulation in 3 redshift bins (open circles: $0.02<z<0.04$; filled circles: $0.04<z<0.10$; filled triangles: $0.1<z<0.2$) without evolution correction. Similar to Fig. 5, we have normalised the LF by Saunders et al.'s estimate of the present-day LF (Eq. 4) to increase the contrast between LFs in different bins. Arp 220-type galaxies exhibit strong negative evolution. For galaxies with significant AGN contributions, there seems to be some indications of strong positive evolution although the error bars are too big for it to be conclusive. Cirrus- and M82-type galaxies show similar weak positive evolutionary trend. Because the redshift range is quite narrow, the effect of evolution is not obvious. In Fig. 9, we show that LF derived from maximum likelihood method assuming a luminosity evolution of $Q=1.7$ (Eq. 13) describes both cirrus and M82 type reasonably well, indicating that the two types have similar evolutionary trends.}
\label{fig:zbin}
\end{figure*}

\subsection{Evolution}

A similar maximum likelihood approach can be used to determine the evolution of the LF (S90; Fisher et al. 1992). For pure density evolution parametrised in Eq.~\ref{densityEvolution}, the evolutionary rate $P$ can be found by maximising the likelihood of observing galaxies at positions $r_1, r_2, ...$ respectively, that is 
\begin{equation}
\prod_i p(z_i|L_i)=\prod_i \frac{(1+z_i)^P dV_i}{\int_{z_{\textrm{min}}}^{z_i^m} (1+z)^P (dV / dz) dz},
\end{equation}
where $z_{\textrm{min}}$ is the lower redshift cutoff.

The above method, known as the constant-density method, is strongly affected by density inhomogeneities in the analysed sample. Therefore, a lower redshift limit is usually used to reduce this effect which is strongest in the local Universe. Table~\ref{evolution} is a compilation of our estimates of $P$ in various redshift bins in an Einstein-de Sitter (to facilitate comparison with others) or $\Lambda$CDM model and previous estimates. Using the constant-density method, S90 detected a strong density evolution $P=6.7\pm2.3$ (Eq. 6). It was confirmed from source number counts in the deep IRAS Faint Source Survey (FSS) data base at $|b|>50^{\circ}$ (Lonsdale et al. 1990). However, Fisher et al. (1992) found no evidence for evolution in their sample of 5,297 IRAS galaxies flux-limited at 1.2 Jy at 60 $\mu m$. Oliver et al. (1995) found $P\approx5.6$ in a spectroscopic sample of around 2,000 galaxies with $F_{60}\ge 0.2$ Jy. Bertin, Dennefeld \& Moshir (1997) also found evidence to support strong evolution in a deep subsample of the IRAS FSS at a flux limit of $F_{60}\approx0.11$ Jy, using several methods such as background fluctuations (i.e. confusion noise), number counts and optical identification rate. SW98 developed a minimum variance estimator which is less affected by density inhomogeneities and re-analysed the 1.2 Jy sample\footnote{The sample used in Fisher et al. (1992) was an early version. The final version was used in the re-analysis by SW98.}. They claimed a milder evolution $P\approx4.3$ in the IRAS galaxies. The IRAS 1 Jy survey of ULIRGs (ultraluminous infrared galaxies with $L_{ir} \ge 10^{12} L_{\odot}$) showed a strong evolution (Kim \& Sanders 1998). Takeuchi, Yoshikawa \& Ishii (2003) applied the constant-density method to the IRAS PSCz and found $P\approx3.4$, consistent with SW98. 

\begin{table*}
\caption{Estimates of the evolutionary rate $P$. The last six rows are estimates derived in this paper.}\label{evolution}
\begin{tabular}[pos]{lll}
\hline
Sample                                              & P        & Method\\
\hline
\hline
QCD survey (S90)                                    & $6.7\pm2.3$  & Constant density method\\
The IRAS Faint Source Survey (Lonsdale et al. 1990) & $\sim7$      & Source number counts\\
1.2 Jy survey (Fisher et al. 1992)                  & $2\pm3$      & Constant density method\\
A deep IRAS redshift survey (Oliver et al. 1995)    & $5.6\pm2.3$  & $\langle V/V_{\textrm{max}} \rangle$, number counts\\
A subsample of the IRAS FSS (Bertin et al. 1997)    & $6.0\pm1.2$  & Background fluctuation analysis, number counts\\  
1.2 Jy survey (SW98)                                &$4.3\pm2.4$   & Minimal variance method\\
IRAS 1 Jy survey of ULIGs (Kim \& Sanders 1998)     &$7.6\pm3.2$   & Constant density method\\
IRAS PSCz (Takeuchi et al. 2003)                    &$3.4\pm0.7$   & Constant density method\\
\textbf{IIFSCz (this paper)}, \boldmath$0.01 \le z \le 0.1$ \textbf{(Einstein-de Sitter)}   &\boldmath$4.7\pm0.9$   & \textbf{Constant density method}\\ 
...................................., \boldmath$0.02 \le z \le 0.1$ \textbf{(Einstein-de Sitter)}   &\boldmath$3.2\pm0.9$   & \textbf{Constant density method}\\
...................................., \boldmath$0.02 \le z \le 0.2$ \textbf{(Einstein-de Sitter)}   &\boldmath$2.3\pm0.6$   & \textbf{Constant density method}\\ 
...................................., \boldmath$0.01 \le z \le 0.1$ (\boldmath$\Lambda$\textbf{CDM})         &\boldmath$4.3\pm0.9$   & \textbf{Constant density method}\\ 
...................................., \boldmath$0.02 \le z \le 0.1$ (\boldmath$\Lambda$\textbf{CDM})         &\boldmath$3.4\pm0.9$   & \textbf{Constant density method}\\
...................................., \boldmath$0.02 \le z \le 0.2$ (\boldmath$\Lambda$\textbf{CDM})         &\boldmath$2.1\pm0.6$   & \textbf{Constant density method}\\ 
\hline
\end{tabular}
\end{table*}

\begin{figure*}\centering
\includegraphics[height=3.0in,width=3.4in]{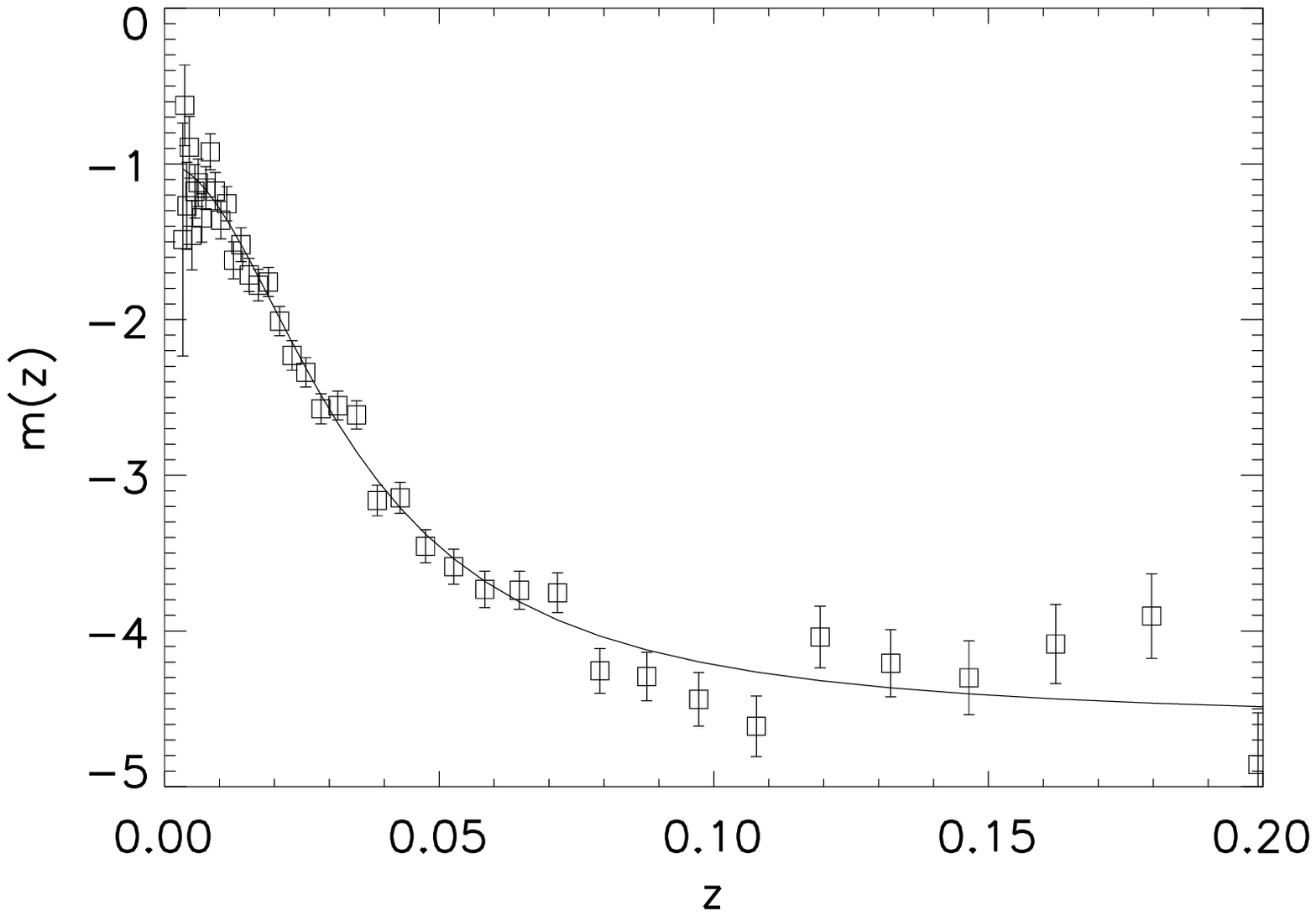}
\includegraphics[height=3.0in,width=3.4in]{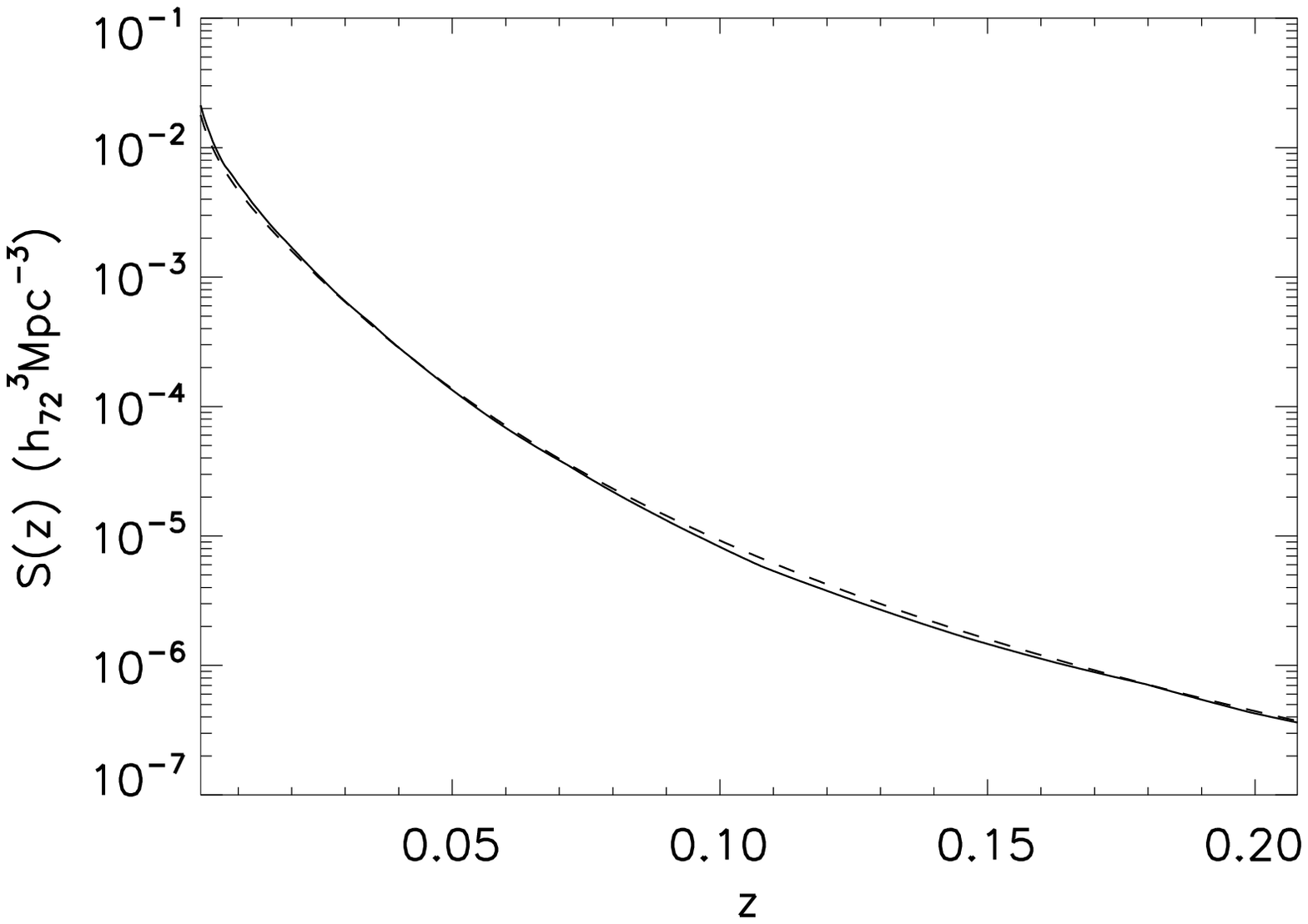}
\caption{Left: The logarithmic slope of the selection function versus redshift. The curve is the best-fit analytic SF as parametrised in Equation 16. Right: The non-parametric SF (solid line) versus the best-fit analytic SF (dashed line). The normalisation is derived from the number density of galaxies with $F_{60}\ge0.36$ Jy per unit solid angle.}
\label{fig:mk}
\end{figure*}

\begin{figure*}\centering
\includegraphics[height=3.0in,width=3.4in]{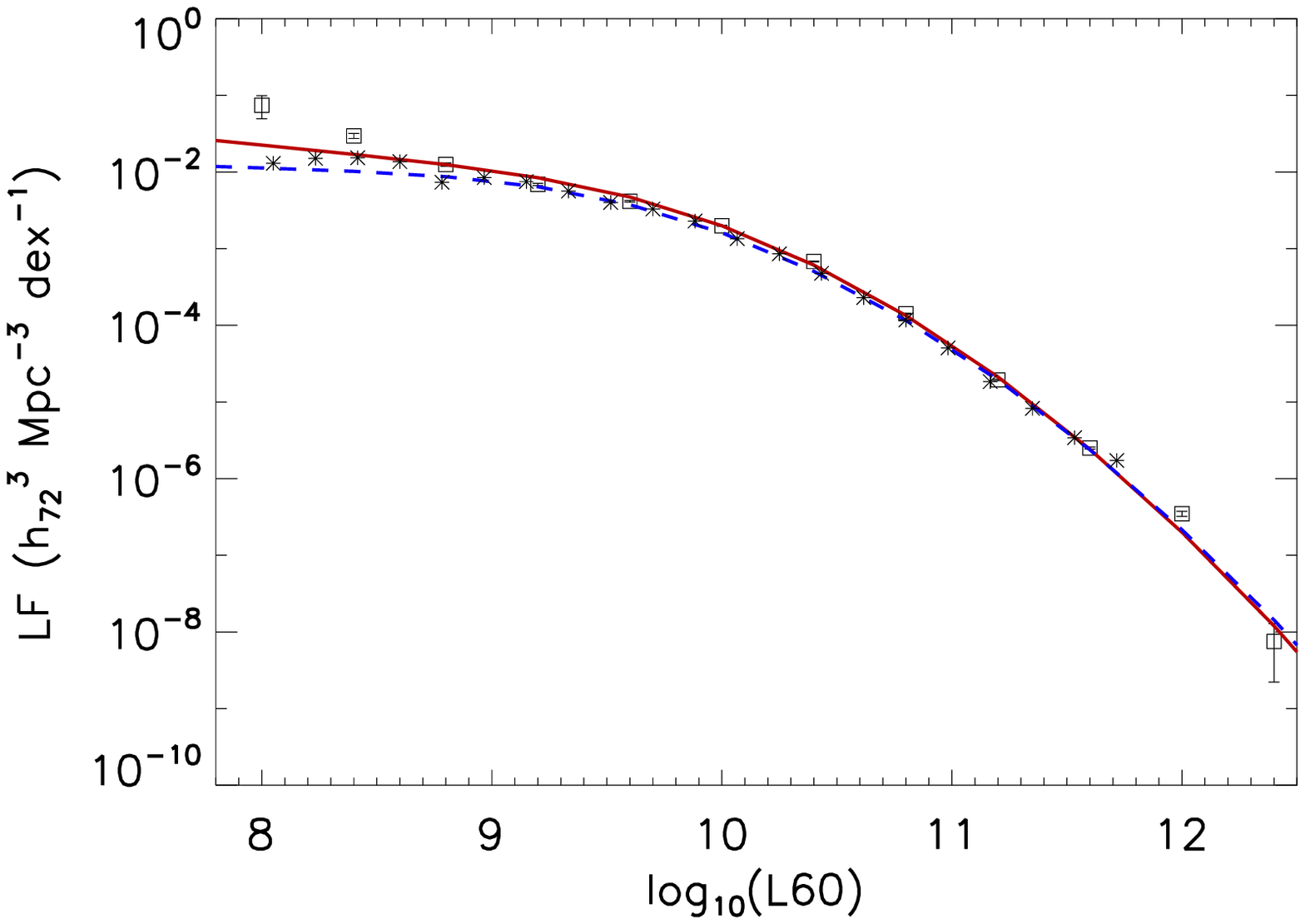}
\includegraphics[height=3.0in,width=3.4in]{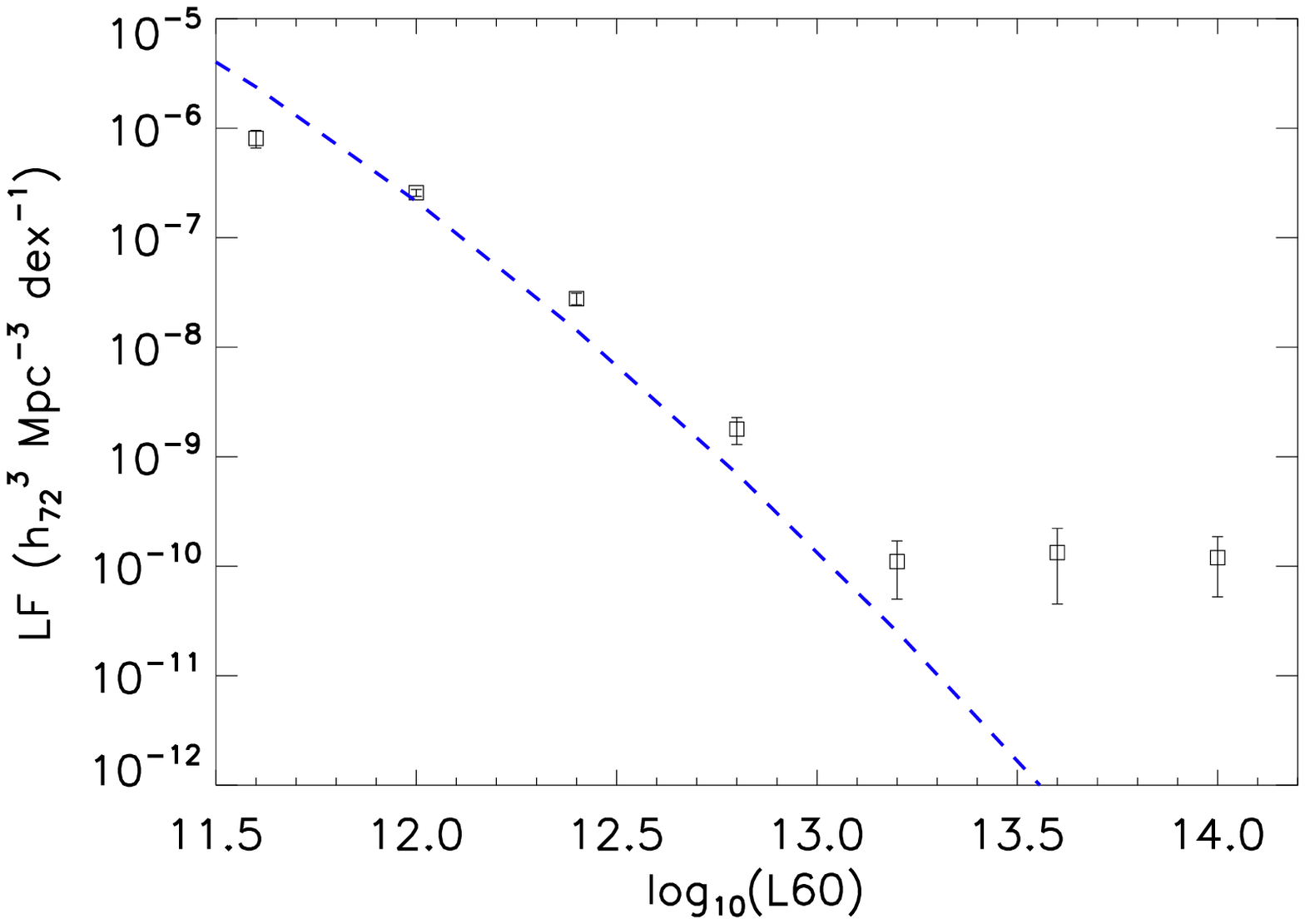}
\caption{Left: The present-epoch LFs in redshift bin [0.003, 0.2] from various estimators after correction for a pure luminosity evolution of $Q=1.7$. Squares - $1/V_{\textrm{max}}$ LF; blue dashed line - estimates from S90; red solid line - our best estimate using the STY method; asterisks -- non-parametric LF. Right: LFs in redshift bin [0.2, 4.0].}
\label{fig:Vmax}
\end{figure*}

\begin{figure*}\centering
\includegraphics[height=4.0in,width=5.0in]{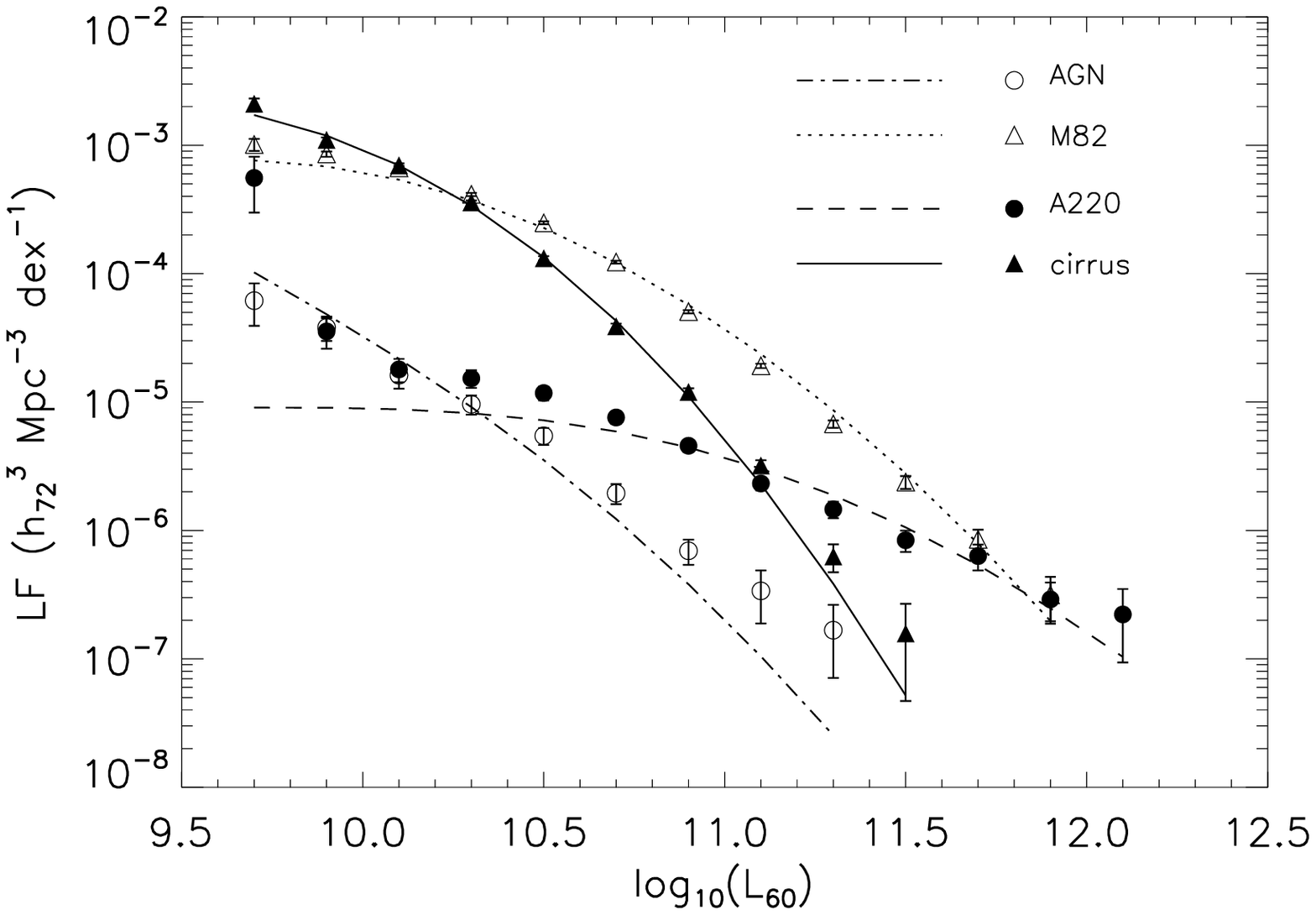}
\caption{Luminosity functions at 60\, $\mu$m derived from the $1/V_{\textrm{max}}$ method (symbols) and the STY method (curves) for cirrus, M82, Arp 220 and AGN dust torus type galaxies in the redshift range $0.02\le z\le0.1$, corrected for a pure luminosity evolution of $Q=1.7$. At $\log_{10}(L_{60}) > 10.3$, M82-type starbursts start to dominate in number until about $\log_{10}(L_{60}) \sim 11.8$. We can also see that the assumed luminosity evolution $Q=1.7$ describes the LF of cirrus- and M82-type galaxies reasonably well. For AGN dust tori, $Q=1.7$ clearly underestimates the true magnitude of evolution, while for Arp 220-type starbursts, the degree of evolution is overestimated.}
\label{fig:LF-nirtem}
\end{figure*}

\subsection{Result and discussion}

Although evolution seems to be detected in most cases, there is disagreement in the literature about the magnitude of evolution. The value of $P$ does not vary much between an Einstein-de Sitter and $\Lambda$CDM model over the limited redshift range examined here. However, it depends quite strongly on the redshift boundaries of the sample. Our adopted value is $P=3.4$ derived from the redshift interval $[0.02, 0.1]$ in which the sample is expected to be relatively complete and less affected by the peculiar velocity field. The pure density evolution can be replaced by pure luminosity evolution with $Q=1.7$ (Eq. 13). In Fig.~\ref{fig:zbin_all}, the $1/V_{\textrm{max}}$ LFs of the full sample are shown in three redshift bins within redshift $z=0.2$. We have corrected for the fact that both the first (faintest) and last (brightest) luminosity bins are usually biased low. Evolution is clearly seen\footnote{If there is positive evolution, higher redshift points appear above the lower redshift points in the luminosity range where they overlap. If there is negative evolution, the reverse is true. E.g. for density evolution, positive evolution means that number density decreases with time (or increases with redshift).} and the magnitude of evolution is consistent with that derived from the maximum likelihood method. 

Galaxy evolution as a function of infrared template type is also investigated. Fig.~\ref{fig:zbin} shows the LFs derived for each infrared galaxy population in three bins in the redshift range [0.02, 0.2]. Apart from Arp 220-type starbursts, all other types seem to experience different degrees of positive evolution\footnote{In Fig. 9, it can be clearly seen that a pure luminosity evolution of magnitude Q=1.7 (derived from the whole sample using maximum likelihood method) fits the evolution in both cirrus and M82 subpopulations quite well. AGN dust tori show stronger evolution than that of cirrus and M82, while Arp 220 - type galaxies have an opposite trend of evolution.}. As a result, the dependence of $P$ on the redshift range is perhaps due to the varying fraction of each infrared subpopulation which undergoes different evolutionary trend. The apparent negative evolution of the A220-type starbursts at low redshift is interesting and perhaps indicates the era of major merger events has ended. However, we do not exclude the possibility that the very different evolutionary trend seen in A220-type starbursts (a very small fraction of the whole infrared population) might be caused by incorrect classification. Presumably A220-type starbursts show strong positive evolution at higher redshift as a result of higher merger rate. Interestingly, this negative evolution at low redshift is predicted in the source count model of Rowan-Robinson (2009). 

The left panel of Fig.~\ref{fig:mk} displays the local logarithmic slope $m_k$ and its errors out to redshift $z=0.2$, corrected for density evolution $P=3.4$. The curve is obtained by a minimum $\chi^2$ fitting with the SF parametrised as, 
\begin{equation}
S(z) = \frac{\psi}{z^\alpha [1+(\frac{z}{z^*})^\gamma]^{\beta/\gamma}},
\end{equation}
where the best-fit parameters are $\alpha=0.999\pm0.076, ~\beta=3.587\pm0.169, ~\gamma=2.001\pm0.165, ~z^*=0.0339\pm0.0014$. The right panel of Fig.~\ref{fig:mk} compares the non-parametric SF with the analytic SF. The normalisation is derived from the number density of galaxies with $F_{60}\geq 0.36$ Jy per unit solid angle ($\sim2567~\textrm{sr}^{-1}$ in the redshift range $[0.003, 0.2]$ in our sample)
\begin{equation}
N = \frac{1}{4\pi} \int S(z) \frac{dV}{dz} dz.
\end{equation} 
From Eq.~\ref{SFdefinition} and Eq.~\ref{SFpowerlaw}, the present-day piecewise LF can be calculated as 
\begin{equation}
\Phi_0 (L) = \left[\frac{g'(z_{\textrm{max}})}{g(z_{\textrm{max}})} -\frac{m_k}{x_k}\right] \left[\frac{z_{\textrm{max}}}{x_k}\right]^{m_k} \times \frac{S_k}{g(z_{\textrm{max}})L'_{\textrm{min}}(z_{\textrm{max}})}
\end{equation}
for $L_{\textrm{min}} (x_{k-1})<L<L_{\textrm{min}}(x_k)$. 

The left panel of Fig.~\ref{fig:Vmax} shows LFs derived from various estimators in the redshift range $0.003\le z \le0.2$, after correction for a pure luminosity evolution of $Q=1.7$. In determining the LF using the $1/V_{\textrm{max}}$ estimator, we weight each galaxy by the redshift completeness as a function of $F_{60}$. The errors are calculated from 
\begin{equation}
\sigma = \sqrt{ \sum_i \frac{\textrm{weight},i}{V_{\textrm{max},i}^2} }.
\end{equation}
The faint end of the $1/V_{\textrm{max}}$ LF is problematic as this method is sensitive to density inhomogeneities which are more pronounced in the very local Universe. The non-parametric LF (derived from the non-parametric SF) shows a good agreement with the estimate from S90 and our estimate using the STY method. Our best-fit parameters in the LF parametrised in Eq.~\ref{eqn:Saunders-LF} are, 
\begin{equation}
\alpha = 1.29, ~L_* = 10^{8.67} h^{-2} L_{\odot}, ~\sigma = 0.72.
\end{equation}
Therefore, our best-estimate LF using the STY method has a steeper faint end than that of S90 and a higher characteristic luminosity $L_*$. The normalisation is obtained via a minimum $\chi^2$ fitting to the $1/V_{\textrm{max}}$ LF (squares in Fig.~\ref{fig:Vmax}),
\begin{equation}
\chi^2 = \sum_i \frac{ (\varphi_i^{\textrm{obs}} - C \varphi_i^{\textrm{STY}})^2}{\sigma_i^2},
\end{equation}
where $\varphi_i^{\textrm{obs}}$ is the measured $1/V_{\textrm{max}}$ LF at luminosity $L_i$ and $\sigma_i$ is the rms error. The normalisation is thus
\begin{equation}
C = \frac{\sum_i \varphi_i^{\textrm{obs}} \varphi_i^{\textrm{STY}} / \sigma_i^2}{\sum_i (\varphi_i^{\textrm{STY}})^2/\sigma_i^2}.
\end{equation}

The right panel of Fig.~\ref{fig:Vmax} shows the $1/V_{\textrm{max}}$ LF in the redshift range $0.2\le z \le 4.0$, corrected for a pure luminosity evolution of $Q=1.7$. Again, it is compared with the present-day LF of S90. The LF begins to flatten at around $L_{60}>10^{13.0} L_{\odot}$, which might be due to significant evolution or a new infrared galaxy population arising at redshift $z>0.2$. We have subsequently obtained 99 hours on ESO VLT to do spectroscopy for these ultraluminous infrared galaxies (ULIRGs) and hyperluminous infrared galaxies (HLIRGs). The nature of these objects will be discussed in future papers.

The present-day LF for each infrared subpopulation is plotted in Fig.~\ref{fig:LF-nirtem}, assuming a common evolution for all template types. Although Fig.~\ref{fig:zbin} suggests different rates of evolution for different spectral types, the samples of the Arp 220-type starbursts and AGN dust tori were not large enough to determine their rate of evolution reliably. we can see that at the low luminosity end, cirrus-type galaxies dominate while M82 starbursts dominate at the high luminosity end (at least up to $10^{12} L_{\odot}$). The subpopulation of AGN and Arp 220-type starbursts exhibit large deviations from the evolution derived for the full sample, reinforcing the conclusion drawn from Fig.~\ref{fig:zbin}.

\subsection{Monte Carlo analysis of photometric redshift errors}

At the flux limit of 0.36 Jy at 60 $\mu$m, less than $15\%$ of the galaxies in the IIFSCz have photometric redshifts obtained through the empirical training set (for galaxies with near-infrared or radio counterparts) or the template-fitting method (for galaxies with optical identifications). Assuming the distribution of photometric redshift is a Gaussian\footnote{It is a strong (and perhaps not very accurate) assumption that photometric redshift errors follow a Gaussian distribution.}, we can use Monte Carlo analysis to investigate the uncertainty in the LF induced by photometric redshift error.

We have created 100 realisations of the flux-limited sample in the redshift range $0.003\le z \le0.2$ with photometric redshifts replaced by random numbers generated from a Gaussian distribution ($\sigma=0.02$) centred at the original photometric redshift estimate. we then derive the $1/V_{\textrm{max}}$ LF for each realisation. The resulting rms scatter is at most $7\%$ (mean scatter $4\%$). Compared to the Poisson error ($45\%$ at most; mean scatter $10\%$), we will neglect the uncertainty induced by the inaccuracy of the photometric redshifts.  

\section{Galaxy bias}
\label{galaxybias}

\subsection{Two-point statistics}

\label{statistics}
The spatial two-point correlation function $\xi(r)$ is one of the most frequently used tools in studying galaxy clustering. It is defined as the probability of finding a galaxy pair at a given separation, in excess of that in a uniform random Poisson distribution. In practise, there are several estimators available:

\begin{itemize}
\item The Blanchard \& Alimi (1988; hereafter BA88) estimator
\begin{equation}
\xi (r) = \frac{DD (r)}{DR (r)} \frac{n_R}{n_D} - 1;
\end{equation}

\item The Landy \& Szalay (1993; hereafter LS93) estimator
\begin{equation}
\xi (r) = \frac{1}{RR(r)}\left[DD(r)(\frac{n_R}{n_D})^2-2DR(r)(\frac{n_R}{n_D})+RR(r)\right];
\end{equation}

\item The Hamilton (1993; hereafter H93) estimator,
\begin{equation}
\xi (r) = \frac{DD(r)RR(r)}{DR^2 (r)} - 1.
\end{equation}

\end{itemize}
$n_D$ and $n_R$ are the mean densities of the galaxy and random catalogues respectively. $DD (r)$, $DR (r)$ and $RR (r)$ are numbers of weighted galaxy-galaxy pairs, galaxy-random pairs and random-random pairs at separation $r$ respectively. For volume-limited samples, the weight function applied to each galaxy is simply $w_i=1$. For flux-limited samples, it is common to use the minimum variance weighting $1/(1+4\pi n(z_i) J_3(r))$, where $n(z_i)$ is the density at $z_i$ and $J_3(r)=\int_0^r \xi(r') r'^2 dr'$. The main difference between these estimators lies in the sensitivity to the uncertainty of the mean density estimation. We found almost identical results from these three methods and will only show correlation functions obtained from the Landy \& Szalay estimator in this paper.

Identical selection effects (both angular and radial) need to be reproduced in creating random catalogues without structures. We use the luminosity functions of cirrus-type and M82-type galaxies to create random samples in calculating the clustering of these two subpopulations. As mentioned in Section 2, the redshift completeness of the IIFSCz varies across the sky. Therefore, we first apply an angular mask which takes into account regions missed by IRAS to the generated random catalogue. Then, we modulate the number of random galaxies in each grid of the sky according to the redshift completeness so that regions with higher redshift completeness will have more random galaxies.

The well known redshift-space distortions consist of small-scale elongation (fingers-of-God effect) and large-scale compression of structures along the line-of-sight. The former is due to random motions in virialized systems while the latter is caused by coherent motions of galaxies towards over-dense regions. To derive the real-space correlation function $\xi(r)$, we calculate the projected two-point correlation function\footnote{The projected correlation function is sometimes denoted as $w_p(r_p)$, where $r_p$ is the separation perpendicular to the line-of-sight.} (Davis \& Peebles 1983)
\begin{equation}
\Xi (\sigma) = 2 \int_0^{\infty} d \pi \xi (\sigma, \pi),
\end{equation}
where $\xi(\sigma, \pi)$ is the two-dimensional correlation function, $\pi$ and $\sigma$ are separations along and perpendicular to the line-of-sight respectively. Thus the bias caused by peculiar motions along the line-of-sight is integrated out. If the real-space two-point correlation function can be well described by a power-law $\xi(r)=(r/r_0)^{-\gamma}$, then
\begin{equation}
\frac{\Xi (\sigma)}{\sigma} = \left(\frac{\sigma}{r_0}\right)^{-\gamma} B\left(\frac{\gamma - 1}{2}, \frac{1}{2}\right).
\end{equation}
Here $B$ is the beta function which satisfies
\begin{equation}
B\left(\frac{\gamma - 1}{2}, \frac{1}{2}\right) = \frac{\Gamma{(\frac{1}{2})} \Gamma{(\frac{\gamma-1}{2})}}{\Gamma(\frac{\gamma}{2})}.
\end{equation}

To quantify the dependence of galaxy clustering on physical parameters such as luminosity and spectral type, we define the relative bias between two classes of galaxies as a function of scale, 
\begin{equation}
\frac{b_1}{b_2}\left(r\right) = \sqrt{\frac{\xi_1 (r)}{\xi_2 (r)}}.
\end{equation}
In this paper, we measure the relative bias of the subsample in the form of 
\begin{equation}
b_{\textrm{rel}}=[\Xi (\sigma)/\Xi_{\textrm{fid}} (\sigma)]^{1/2},
\label{eqn:bias}
\end{equation}
where the assumed fiducial projected correlation function corresponds to a real-space correlation function $\xi (r) = (r/3.5)^{-1.8}$. The same approach was used in Zehavi et al. (2005).

\begin{table*}\centering
\caption{Nested volume-limited samples of all galaxies (from V1 to V7), M82-type starbursts (from MV1 to MV6) and cirrus-type quiescent galaxies (from CV1 to CV5). The columns are sample name, redshift range, 60 micron luminosity threshold, mean 60 $\mu$m luminosity, number of galaxies, correlation length $r_0$, power-law index $\gamma$, mean bolometric infrared luminosity derived from the best-fit infrared template and SFR.}\label{VLsubsamples}
\begin{tabular}[pos]{lllllllll}
\hline
Sample  & z range      &$\textrm{Min}(\log(L_{60}))$&$\langle \log(L_{60})\rangle$&$N_g$&$r_0$ ($h_{72}^{-1}$ Mpc)&$\gamma$&$\langle \log(L_{IR}) \rangle$&$\langle \textrm{SFR} \rangle$ $(M_{\odot} /yr)$\\
\hline           
\hline                              
V1  &[0.003, 0.01] &9.00 & $9.48\pm0.38$  & 1010  &$3.75\pm0.46$&$2.55\pm0.24$&$9.91\pm0.37$ &1.00\\
V2  &[0.003, 0.02] &9.60 & $9.94\pm0.28$  & 2352  &$3.30\pm0.21$&$1.67\pm0.06$&$10.33\pm0.29$&2.87\\
V3  &[0.003, 0.03] &9.96 & $10.24\pm0.24$ & 3081  &$2.85\pm0.24$&$1.69\pm0.07$&$10.55\pm0.26$&5.71\\
V4  &[0.003, 0.04] &10.22& $10.46\pm0.22$ & 3283  &$3.42\pm0.35$&$1.74\pm0.09$&$10.79\pm0.29$&9.56\\
V5  &[0.003, 0.05] &10.42& $10.64\pm0.21$ & 2916  &$3.50\pm0.43$&$1.66\pm0.10$&$11.00\pm0.25$&14.50\\
V6  &[0.003, 0.06] &10.58& $10.80\pm0.20$ & 2617  &$3.05\pm0.45$&$1.60\pm0.16$&$11.10\pm0.24$&20.76\\
V7  &[0.003, 0.07] &10.72& $10.92\pm0.19$ & 2349  &$3.03\pm1.08$&$1.83\pm0.37$&$11.26\pm0.23$&27.68\\
\end{tabular}
\begin{tabular}[pos]{lllllllll}
\hline
Sample    & z range      &$\textrm{Min}(\log(L_{60}))$&$\langle \log(L_{60})\rangle$&$N_g$&$r_0$ ($h_{72}^{-1}$ Mpc)&$\gamma$&$\langle \log(L_{IR}) \rangle$&$\langle \textrm{SFR} \rangle$ $(M_{\odot} /yr)$\\
\hline
\hline
MV1  &[0.003, 0.01] &9.00 & $9.61\pm0.44$  & 343   &$3.39\pm1.20$ &$1.89\pm0.42$&$9.97\pm0.44$ &1.36\\
MV2  &[0.003, 0.02] &9.60 & $10.01\pm0.31$ & 1050  &$2.75\pm0.39$ &$1.63\pm0.14$&$10.36\pm0.30$&3.40\\
MV3  &[0.003, 0.03] &9.96 & $10.29\pm0.26$ & 1650  &$2.60\pm0.83$ &$1.64\pm0.24$&$10.57\pm0.26$&6.42\\
MV4  &[0.003, 0.04] &10.22& $10.49\pm0.23$ & 1984  &$3.54\pm0.46$ &$1.89\pm0.16$&$10.80\pm0.30$&10.30\\
MV5  &[0.003, 0.05] &10.42& $10.66\pm0.22$ & 1990  &$3.38\pm0.87$ &$1.64\pm0.20$&$11.01\pm0.25$&15.21\\
MV6  &[0.003, 0.06] &10.58& $10.81\pm0.20$ & 1880  &$3.99\pm1.08$ &$2.01\pm0.36$&$11.10\pm0.23$&21.13\\
\end{tabular}
\begin{tabular}[pos]{lllllllll}
\hline
Sample    & z range      &$\textrm{Min}(\log(L_{60}))$ &$\langle \log(L_{60})\rangle$&$N_g$&$r_0$ ($h_{72}^{-1}$ Mpc)&$\gamma$&$\langle \log(L_{IR})\rangle$&$\langle \textrm{SFR} \rangle$ $(M_{\odot} /yr)$\\
\hline
\hline
CV1  &[0.003, 0.01] &9.00 & $9.42\pm0.32$  & 624   &$3.56\pm0.43$ &$2.41\pm0.50$&$9.89\pm0.32$ &0.86\\
CV2  &[0.003, 0.02] &9.60 & $9.88\pm0.23$  & 1158  &$4.27\pm0.20$ &$1.87\pm0.06$&$10.34\pm0.22$&2.52\\
CV3  &[0.003, 0.03] &9.96 & $10.17\pm0.18$ & 1157  &$2.41\pm1.01$ &$1.51\pm0.22$&$10.56\pm0.19$&4.93\\
CV4  &[0.003, 0.04] &10.22& $10.39\pm0.15$ & 972   &$4.31\pm1.12$ &$1.84\pm0.26$&$10.80\pm0.25$&8.12\\
CV5  &[0.003, 0.05] &10.42& $10.56\pm0.12$ & 580   &$3.47\pm2.16$ &$1.69\pm1.02$&$11.01\pm0.21$&11.99\\
\hline
\end{tabular}
\end{table*}

\subsection{Error analysis}

Due to the correlated nature of the correlation function at different separations, Poisson error may seriously underestimate the true uncertainties in the correlation function. To determine the covariance of our data, we use the bootstrap resampling technique (Ling, Frenk \& Barrow 1986) to create a series of data sets. 

\begin{equation}
\Delta(s_i) = \frac{\xi(s_i) - \overline{\xi(s_i)}}{\sigma(s_i)},
\end{equation}
where $\overline{\xi(s_i)}$ is the mean value of the correlation function at separation $s_i$ from the different bootstrap samples and $\sigma(s_i)$ is the rms scatter. The covariance error matrix is 
\begin{equation}
C_{ij} = \langle \Delta(s_i) \Delta(s_j) \rangle.
\end{equation}
The maximum-likelihood estimate of the parameters in the model can be calculated by minimising
\begin{equation}
\chi^2 = (\Delta^{\textrm{res}})^T C^{-1} \Delta^{\textrm{res}},
\end{equation}
where $\Delta^{\textrm{res}} = ( \xi^{\textrm{observed}} - \xi^{\textrm{model}} )/\sigma$. However, the bootstrap resampling method does not account for uncertainties caused by fluctuations outside the sampled volume. In all correlation functions presented below, we have used the full covariance matrix to derive the best-fitting power law. 

\begin{figure*}\centering
\includegraphics[height=4.5in,width=6.5in]{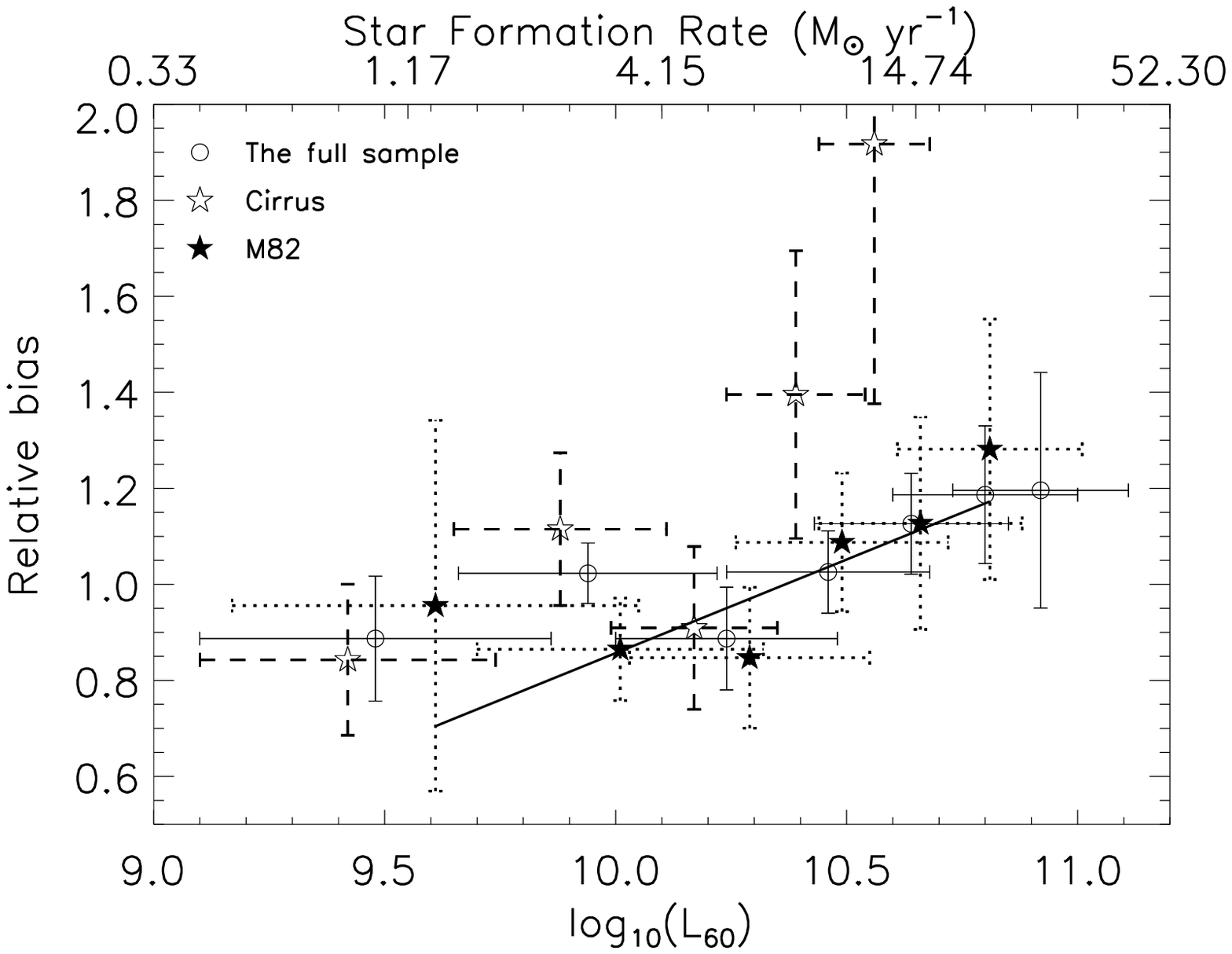}
\caption{Relative bias for volume-limited samples listed in Table 2 as a function of 60 micron luminosity. The relative bias is calculated using Eq.~\ref{eqn:bias} and $\sigma=3$ Mpc. The error bars are derived from 25 bootstrap realisations of each subsample. At $\log_{10}(L_{60}) < 10.3$, cirrus-type galaxies dominate in number (see Fig. 9), so galaxy clustering of the whole sample (open circles) follows that of cirrus-type (empty stars). At higher luminosities $\log_{10}(L_{60}) > 10.3$, M82-type starbursts start to dominate and the clustering of the full sample is almost the same as that of M82-type starbursts. The thick solid line is the best-fit straight line of the luminosity-dependence of clustering for M82-type starbursts which clearly exhibits a correlation between SFR and clustering strength.}
\label{fig:bias_luminosity}
\end{figure*}

\subsection{The luminosity dependence of clustering}

The infrared emission of galaxies comes from dust emission associated with regions of active star formation as well as the general interstellar radiation field. Luminous infrared galaxies are often starbursts or dust-enshrouded AGN. As a result, the FIR luminosity is expected to be an excellent SFR tracer for strong compact dusty starbursts, but is less reliable for the disks of normal galaxies where radiation from old stars contribute significantly to the heating of dust (Helou 1986; Kennicutt 1988; Rowan-Robinson et al. 1997). In this paper, we use the 60 micron luminosity to derive SFR (Rowan-Robinson et al. 1997),
\begin{equation}
\phi_* (M_{\odot}~\textrm{yr}^{-1}) = 2.2~\epsilon^{-1}~10^{-10}~\left(\frac{L_{60}}{L_{\odot}}\right),
\label{eqn:SFR}
\end{equation}
where $\epsilon=2/3$ is the adopted fraction of UV light absorbed by dust. 

Previous studies using smaller samples were unable to detect any luminosity-dependence of the clustering amplitude, possibly due to the contrasting effects of the dark matter halo mass and specific star formation rate (SSFR) on determining the far-infrared luminosity\footnote{The SSFR is lower in more massive galaxies at both the local Universe $z\sim0$ and the distant Universe $z\sim1$ (Brinchmann et al. 2004; Elbaz et al. 2007; Noeske et al. 2007; Zheng et al. 2007; Damen et al. 2009). Oliver et al. (in preparation) use the Spitzer Wide-area Infrared Extragalactic Legacy Survey (SWIRE) to study the SSFR as a function of stellar mass parametrised as SSFR $\propto M_{\ast}^{\beta}$, where $M_{\ast}$ is stellar mass. They find that the SSFR shows a less steep decline for late-type galaxies ($\beta \sim -0.15$) than early type galaxies ($\beta \sim -0.46$).}. Based upon the luminosity - colour correlation and the colour dependence of the correlation function, Hawkins et al. (2001) proposed a luminosity-dependence of the form
\begin{equation} 
\log_{10}(s_0) = -0.028 \log_{10}(L_{60}) + 0.95,
\end{equation}
where $s_0$ is the correlation length in the redshift space. So, unlike optical galaxies, they found a weaker correlation for infrared galaxies of higher luminosity but their result is of low statistical significance. However, there might be a problem with their estimate because they used the redshift-space correlation function. The effect of redshift distortion at the scale we are interested in tends to boost the clustering signal and it is worse for nearby samples where peculiar velocities along the line-of-sight are not negligible. Thus, the correlation length derived from redshift-space correlation function will be higher than the true value for nearby samples. It is also important to bear in mind that Eq. 35 was proposed to describe the luminosity-dependence for the full sample including both cirrus and M82 type. Therefore, it is not really relevant to the question we are trying to address in this paper, i.e. the relation between SFR and environment for star-forming galaxies. 

To investigate the luminosity dependence of galaxy clustering and at what range it occurs, the IIFSCz is divided into nested volume-limited subsamples. In Table~\ref{VLsubsamples}, we have listed redshift range, 60 $\mu$m luminosity threshold, mean 60 $\mu$m luminosity, number of galaxies in the sample, correlation length $r_0$ in the real-space, power-law index $\gamma$, the bolometric infrared luminosity derived from the best-fit infrared template and SFR derived from Eq. 34. We also extract volume-limited samples of cirrus-type galaxies and M82-type starbursts. In Fig.~\ref{fig:bias_luminosity}, the relative bias defined by Eq. 30 with $\sigma=3$ Mpc is plotted against the mean 60 micron luminosity $L_{60}$ for each subsample in Table~\ref{VLsubsamples}. The error bars are the rms scatter derived from 25 bootstrap realisations. The empty stars and the filled stars represent the relative bias for cirrus and M82 subsamples respectively. The open circles show the relative bias of the full sample, which closely follow the relative bias of cirrus-type galaxies at low luminosities $\log_{10}(L_{60}) < 10.3$ and then M82-type starbursts at higher luminosities. This behaviour of the whole sample is expected as we have already seen from Fig. 3 and Fig. 9 that cirrus-type galaxies is the dominant subpopulation at low redshift / luminosity and then M82-type starbursts take over at around $\log_{10}(L_{60}) \approx 10.3$. 

For cirrus-type quiescent galaxies, dust emission caused by old stellar population contributes significantly to the infrared emission and thus the derived SFR are subject to large uncertainties. To study the relation between current star formation activity and environment, it is more relevant to look at the clustering of M82-type starbursts as a function of infrared luminosity. In Fig.~\ref{fig:bias_luminosity}, we can see that the correlation strength correlates tightly with SFR for M82-type starbursts. The thick solid line is the best-fitting straight line of the SFR - clustering strength correlation. Possible cause for this tight correlation will be discussed in Section 4.5 where our result is compared with clustering of star-forming galaxies in the distance Universe (at $z\sim1$).

\begin{figure*}\centering
\includegraphics[height=5.1in,width=7.2in]{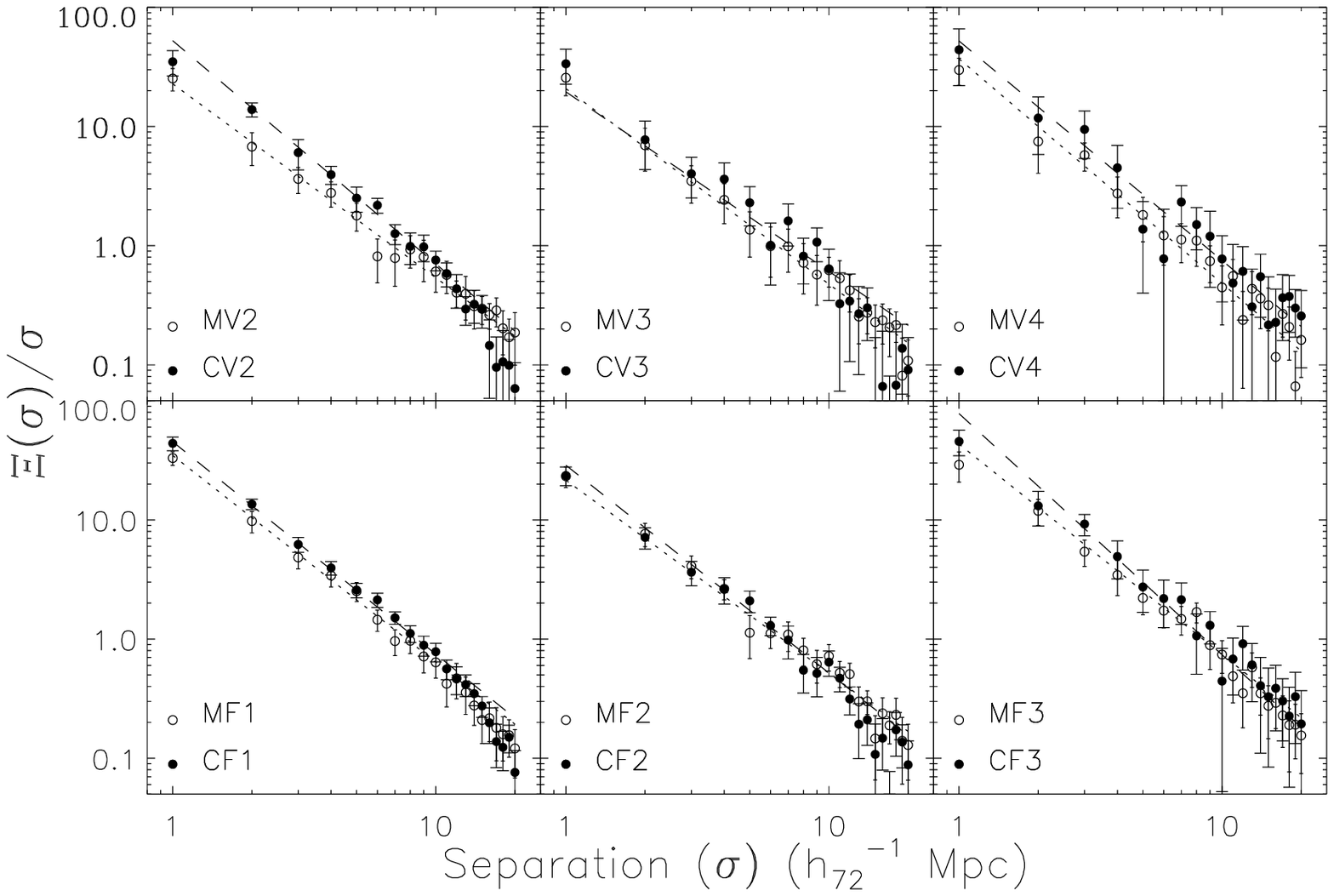}
\caption{Top row: The projected two-point correlation function for selected volume-limited samples of cirrus-type (filled circles) and M82-type (open circles) galaxies listed in Table 2. The dashed (dotted) line is the best-fitting power law to the cirrus (M82) correlation function on scales $2\le\sigma\le13$ Mpc, using the full covariance matrix. The error bars represent the rms spread derived from 25 bootstrap realisations. Bottom row: The projected two-point correlation function for selected flux-limited samples of cirrus-type and M82-type galaxies listed in Table 3.}
\label{fig:wp_cirrus-M82}
\end{figure*}

\subsection{The infrared template type dependence of clustering}

\begin{table*}\centering
\caption{Flux-limited subsamples of M82-type starbursts (from MF1 to MF3) and cirrus-type galaxies (from CF1 to CF3). The columns are sample name, redshift range, mean 60 $\mu$m luminosity, number of galaxies, correlation length and the power-law slope.}\label{FLsubsamples}
\begin{tabular}[pos]{llllll}
\hline
Sample    & z range       &$\langle \log(L_{60}) \rangle$&$N_g$&$r_0$ ($h_{72}^{-1}$ Mpc)&$\gamma$\\
\hline
\hline
MF1  &[0.003,0.015] & $9.52\pm0.55$  & 1027  &$3.52\pm0.31$ &$1.73\pm0.11$\\
MF2  &[0.015,0.03]  & $10.11\pm0.33$ & 2205  &$2.65\pm0.50$ &$1.62\pm0.13$\\
MF3  &[0.03,0.05]   & $10.49\pm0.27$ & 2777  &$3.96\pm0.30$ &$1.76\pm0.10$\\
\end{tabular}
\begin{tabular}[pos]{llllll}
\hline
Sample    & z range       &$\langle \log(L_{60})\rangle$&$N_g$&$r_0$ ($h_{72}^{-1}$ Mpc)&$\gamma$\\
\hline
\hline
CF1  &[0.003,0.015] & $9.34\pm0.44$  & 1710  &$4.05\pm0.20$ &$1.78\pm0.05$\\
CF2  &[0.015,0.03]  & $9.95\pm0.26$  & 2225  &$3.15\pm0.23$ &$1.74\pm0.07$\\
CF3  &[0.03,0.05]   & $10.35\pm0.19$ & 1440  &$4.93\pm0.31$ &$2.02\pm0.17$\\
\hline
\end{tabular}
\end{table*}

The top panel of Fig. 1 in Wang \& Rowan-Robinson (2008) shows the ratio of $F_{100}$ and $F_{60}$ as a function of spectroscopic redshift compared with predictions based on four infrared templates. At redshift $z<1.2$, cirrus-type quiescent galaxies generally have cooler FIR colours than M82 starbursts. In Hawkins et al. (2001), cold galaxies are defined by $F_{100}/F_{60} > 2.3$ which are essentially cirrus-type galaxies in our sample. The colour-luminosity correlation means galaxies of different colour will preferentially be found in different volumes. To reduce the effect of cosmic variance and luminosity dependence of clustering (see the previous section), we need to compare subsamples of cirrus-type galaxies and M82 starbursts, corresponding to cooler and hotter galaxies respectively, in the same volume (redshift range) and same luminosity bin. 

Already in Fig.10, we can see that cirrus-type galaxies have larger relative bias than M82-type starburst in all luminosity bins. Fig.~\ref{fig:wp_cirrus-M82} compares the projected two-point correlations functions of the two infrared subpopulations. The top row shows $\Xi (\sigma) / \sigma$ for selected volume-limited samples of the two types listed in Table 2. The selection is based on the number of galaxies to ensure that we are not comparing very noisy correlation functions. Similarly, the bottom row shows the $\Xi (\sigma) / \sigma$ for selected flux-limited samples of the two types listed in Table 3. Difference in clustering strength between cirrus and M82 is clearly present in the sense that cirrus-type galaxies are more strongly clustered than M82 starbursts. We can also derive the correlation lengths with fixed power-law index $\gamma=1.8$ for subsamples of cirrus and M82-type galaxies listed in Table 2 and the relative bias is calculated to be $b_{\textrm{cirrus}} / b_{\textrm{M82}}=1.25\pm 0.07$ on scales $2\le r\le13~h_{72}^{-1}$ Mpc. At scales larger than $\sim$10 Mpc, there is some indication that the relative bias between the two infrared types decreases. Our result is consistent with the conclusions of Hawkins et al. (2001) that cooler galaxies cluster more strongly than hotter galaxies.  

The clustering dependence on infrared template type is perhaps analogous to the colour (or optical spectral type) dependence found in optical galaxies (Madgwick et al. 2003; Zehavi et al. 2002, 2005; Wang et al. 2007; Cresswell \& Percival 2009; Ma et al. 2009). It suggests that the bulk of infrared luminosity for the two types, cirrus and M82, arises from different sources. The FIR emission of cirrus-type galaxies could be mainly due to dust heated by the general interstellar radiation field where old stars dominate. Given the local SFR - density relation, star-forming galaxies are preferentially found in low-density regions and so it is expected that the clustering strength of M82-like starbursts is weaker than that of cirrus-type quiescent galaxies.

\subsection{Comparison with the distant Universe ($z\sim1$)}
At $z\sim1$ , Cooper et al. (2006) found luminous blue galaxies in the DEEP2 Galaxy Redshift Survey are found in environments with greater overdensity. There are also evidence that clustering strength of star-forming galaxies correlates with SFR in the distant Universe. Gilli et al. (2007) used star-forming galaxies detected at 24 micron by Spitzer/MIPS in the Great Observatories Origins Deep Survey (GOODS) fields to measure the spatial clustering. They found galaxies of higher infrared luminosity have a larger correlation length than less luminous ones. It indicates that more intensive star formation is hosted by more massive dark matter halos. 

The stronger clustering of star-forming galaxies with higher SFR detected in the distant Universe can be explained by the tight correlation between SFR and stellar mass (Noeske et al. 2007; Elbaz et al. 2007), assuming more massive galaxies generally reside in more massive dark matter halos. However, because the SFR - stellar mass relation is also valid in the local Universe, we should expect a similar correlation between clustering strength and SFR which is exactly what we see in the clustering dependence on infrared luminosity for M82-type starbursts (the solid line in Fig. 10).

\section{Discussion and conclusion}
\label{discussions and conclusions}

We have derived the luminosity function and selection function of the IIFSCz to learn about the evolution of infrared galaxies. To minimize the effect of incompleteness, we chose a flux limit of 0.36 Jy at 60 $\mu$m. Either a pure density evolution $P=3.4$ or pure luminosity evolution $Q=1.7$, is detected in the full sample using the maximum likelihood method. The magnitude of evolution is at the lower end of previous estimates. The value of $P$ shows some dependence on the redshift boundaries which might be caused by different mixtures of infrared subpopulations which have very different evolutionary trends. The present-epoch 60 $\mu$m luminosity function derived from galaxies in the redshift range $0.003<z<0.2$ shows good agreement with the luminosity function from Saunders et al. (1990). However, the luminosity function derived from galaxies in the redshift range $0.2<z<4.0$ shows evidence for much stronger evolution or possibly an entirely new class of infrared galaxies of high luminosity emerging at redshift $z>0.2$. Follow-up spectroscopy studies will investigate further the population at the high luminosity end.

The spatial clustering of galaxies as a function of galaxy properties such as type and luminosity provides strong constraints on models of galaxy formation and evolution. In particular, the clustering of the infrared galaxies contains essential information on the relation between star formation activity and environment. We use the projected correlation function $\Xi(\sigma)$, which is free from redshift-space distortion, to study the clustering dependence on infrared template type and infrared luminosity. When comparing clustering strength between the two infrared subpopulations, we found that cirrus-type galaxies cluster more strongly than M82 starbursts in all luminosity bins until the cirrus-population becomes absent at high luminosity/redshift. The average relative bias between the two infrared populations is $b_{\textrm{cirrus}}/b_{\textrm{M82}}=1.25\pm0.07$ on scales $2\le r \le13~h_{72}^{-1}$ Mpc. This is consistent with the SFR - density relation which states that star-forming galaxies preferentially reside in low density environments. The dependence of clustering strength on SFR (using infrared luminosity as a proxy) is investigated using M82-type starbursts. We detect a correlation between SFR and clustering strength, which follows the basic trend found in the distant Universe (at $z\sim1$). It can probably be explained by the tight correlation between SFR and stellar mass which holds at the local Universe as well as $z\sim1$ and possibly $z\sim2$. However, the trend that the SFR increases with clustering strength must becomes invalid at some point, as we do not find galaxies with the highest SFR in environments of highest density (such as centres of rich clusters) in the local Universe. The two seemingly inconsistent statements, the SFR - density relation and the correlation between SFR and clustering strength, can be reconciled if the latter operates mainly in the low density regime and becomes invalid once the density exceeds a certain threshold.

Direct measurement of the infrared emission of galaxies is crucial to measuring SFR of distant galaxies. We hope to extend our study of the relation between SFR and environment to higher redshift using surveys such as SWIRE and COSMOS. Observations to be made by the Herschel satellite will undoubtedly provide more insight. 

\section*{ACKNOWLEDGEMENTS}
We wish to thank the referee for comments which helped to improve the quality of the paper. L. Wang acknowledges Seb Oliver, Isaac Roseboom, David Elbaz and Ho Seong Hwang for helpful discussions. L. Wang has benefited from funding from STFC.

\end{document}